\documentclass[11pt]{cernrep}
\usepackage{cite,./mcite}
\usepackage{epsfig}

\def\p{I\!\!P}

\def\beq{\begin{equation}}
\def\eeq{\end{equation}}
\def\bea{\begin{eqnarray}}
\def\eea{\end{eqnarray}}
\begin{document}
\title{Survival probability in diffractive dijet photoproduction}
\author{Michael Klasen$^1$ and Gustav Kramer$^2$}
\institute{$^1$ Laboratoire de Physique Subatomique et de Cosmologie,
 Universit\'e Joseph Fourier / CNRS-IN2P3 / INPG, 53 Avenue des Martyrs,
 F-38026 Grenoble, France \\
 $^2$ {II.} Institut f\"ur Theoretische Physik, Universit\"at
 Hamburg, Luruper Chaussee 149, D-22761 Hamburg, Germany}
\maketitle
\begin{abstract}
We confront the latest H1 and ZEUS data on diffractive dijet photoproduction with
next-to-leading order QCD predictions in order to determine whether a rapidity gap
survival probability of less than one is supported by the data. We find evidence
for this hypothesis when assuming global factorization breaking for both the
direct and resolved photon contributions, in which case the survival probability
would have to be $E_T^{jet}$-dependent, and for the resolved or in addition the
related direct initial-state singular contribution only, where it would be
independent of $E_T^{jet}$.
\end{abstract}

%%%%%%%%%%%%%% Begin Main Part %%%%%%%%%%%%%%%%%%%%%%%%%%%%%%%%%%%%%%%%%

\vspace*{-100mm}
\noindent LPSC 08-115\\
\vspace*{85mm}

\section{Introduction}
The central problem for hard diffractive scattering processes, characterized by a
large rapidity gap in high-energy collisions, is whether they can be factorized
into non-perturbative diffractive parton density functions (PDFs) of a colorless
object (e.g.\ a pomeron) and perturbatively calculable partonic cross sections.
This concept is believed to hold for the scattering of point-like electromagnetic
probes off a hadronic target, such as deep-inelastic scattering (DIS) or direct
photoproduction \cite{Collins:1997sr}, but has been shown to fail for purely
hadronic collisions \cite{Collins:1997sr,Affolder:2000vb}. Factorization is thus
expected to fail also in resolved photoproduction, where the photon first
dissolves into partonic constituents, before these scatter off the hadronic
target. The separation of the two types of photoproduction processes is, however,
a leading order (LO) concept. At next-to-leading order (NLO) of perturbative QCD,
they are closely connected by an initial-state (IS) singularity originating from
the splitting $\gamma \to q\bar{q}$ (for a review see \cite{Klasen:2002xb}),
which may play a role in the way factorization breaks down in diffractive 
photoproduction \cite{Klasen:2005dq}. The breaking of the resolved
photoproduction component only leads to a dependence of the predicted cross
section on the factorization scale $M_{\gamma}$ \cite{Klasen:2005dq}. Since this
$M_{\gamma}$-dependence is unphysical, it must be remedied also for the
factorization breaking of the resolved part of the cross section, e.g.\ by
modifying the IS singular direct part. A proposal how
to achieve this has been worked out in \cite{Klasen:2005dq} and has been reviewed
already in the proceedings of the workshop on {\em HERA and the LHC} of 2004-2005
\cite{Bruni:2005eb} (see also \cite{Klasen:2005cz}). Since from a theoretical
point of view only the suppression of the resolved or in addition the IS singular
direct component \cite{Klasen:2005dq} is viable, it is an interesting question
whether the diffractive dijet photoproduction data show breaking of the
factorization, how large the suppression in comparison to no breaking will be,
and whether the breaking occurs in all components or just in the resolved plus
direct IS component. The value of the suppression factor or survival probability
can then be compared to theoretical predictions \cite{Kaidalov:2003xf}
and to the survival probability observed in jet production in
$p\bar{p}$ collisions at the Tevatron \cite{Affolder:2000vb} and will be of
interest for similar diffractive processes at the LHC.

Since 2005 no further developments occurred on the theoretical side. On the
experimental side, however, the final diffractive PDFs (DPDFs), which have
been determined from the inclusive measurements of the diffractive structure 
function $F_2^D$ by the H1 collaboration, have been published \cite{Aktas:2006hy}.
Also both collaborations at HERA, H1 and ZEUS, have now published their
final experimental data of the cross sections for diffractive dijet
photoproduction \cite{Aktas:2007hn,Chekanov:2007rh}. Whereas H1 confirm in \cite{Aktas:2007hn} their earlier
findings based on the analysis of preliminary data and preliminary DPDFs,
the authors of the ZEUS analysis \cite{Chekanov:2007rh} reached somewhat different
conclusions from their analysis. Specifically, the H1 collaboration
\cite{Aktas:2007hn} obtained a global suppression of their measured cross sections as
compared to the NLO calculations. In this comparison \cite{Aktas:2007hn},
the survival probability is 
R = 0.5, independent of the DPDFs fit used, i.e.\ fit A or B in Ref.\
\cite{Aktas:2006hy}. In 
addition they concluded that the assumption that the direct cross section obeys
factorization is strongly disfavored by their analysis. The ZEUS collaboration,
on the other hand, concluded from their analysis \cite{Chekanov:2007rh}, that, within the 
large uncertainties of the NLO calculations, their data are compatible with the
QCD calculations, i.e.\ that no suppression would be present.

Due to these somewhat inconsistent results we made a new effort \cite{Klasen:2008ah} to
analyze the H1 \cite{Aktas:2007hn} and the ZEUS \cite{Chekanov:2007rh} data, following more or
less the same strategy as in our earlier work \cite{Klasen:2004tza,Klasen:2004qr} on the basis of
the NLO program of \cite{Klasen:2004tza,Klasen:2004qr} and the new DPDFs sets of Ref.\ \cite{Aktas:2006hy}. The H1
and the ZEUS dijet data cannot be compared directly, since they have
different kinematic cuts. In particular, in the H1 measurements \cite{Aktas:2007hn}
$E_T^{jet1(2)} > 5$ (4) GeV and  $x_{\p} < 0.03$, and in the ZEUS measurements 
\cite{Chekanov:2007rh} $E_T^{jet1(2}) >  7.5$ (6.5) GeV and $x_{\p} < 0.025$ 
(these and all other variables used in this review are defined in
\cite{Klasen:2008ah,Klasen:2004tza,Klasen:2004qr} and in the corresponding
experimental contribution in these proceedings).
It is clear that in order to establish a global suppression, i.e.\ an
equal suppression of the direct and the resolved cross section, the
absolute  normalization and not so much the shape of the measured cross
section is very important. This normalization depends on the applied
kinematic cuts. Of course, the same cuts must be applied to the NLO
cross section calculation. In case of a resolved suppression only, the 
suppression depends
on the normalization of the cross sections, but also on the shape of
some (in particular the $x_{\gamma}^{obs}$, $E_T^{jet1}$, $M_{12}$, and
$\bar{\eta}^{jets}$) distributions,
and will automatically be smaller at large $E_T^{jet1}$ \cite{Klasen:2002xb}.
Distributions in $x_{\p}$ and $y$ (or $W$) are not sensitive to the
suppression mechanism. The distribution in $z_{\p}$, on the other hand, is
sensitive to the functional behavior of the DPDFs, in particular of the
gluon at large fractional momenta.

In the meantime, the H1 collaboration made an effort to put more light into the
somewhat contradictory conclusions of the H1 \cite{Aktas:2007hn} and ZEUS
\cite{Chekanov:2007rh} collaborations by performing a new analysis of their data,
now with increased luminosity, with the same kinematic cuts as in
\cite{Aktas:2007hn}, i.e.\ the low-$E_T^{jet1}$ cut, and the high-$E_T^{jet1}$
cut as in the ZEUS analysis \cite{Chekanov:2007rh}. The results have been
presented at DIS 2008 \cite{h1dis08} and will be published soon. We have
performed a new study of these H1 \cite{h1dis08} and ZEUS data
\cite{Chekanov:2007rh} to show more clearly the differences between the three
data sets \cite{kk08tbp}. In this contribution we shall show a selection of
these comparisons. The emphasis in these comparisons will be, how large the
survival probability of the diffractive dijet cross section will be globally
and whether the model with resolved suppression only will also describe
the data in a satisfactory way. In section 2 we show the comparison with the H1
data \cite{h1dis08} and in section 3 with the ZEUS data \cite{Chekanov:2007rh}.
Section 4 contains our conclusions.

\section{Comparison with recent H1 data}
The recent H1 data for diffractive photoproduction of dijets \cite{h1dis08} have
several advantages as compared to the earlier H1 \cite{Aktas:2007hn} and ZEUS
\cite{Chekanov:2007rh} analyses. First, the integrated luminosity is three times higher
than in the previous H1 analysis \cite{Aktas:2007hn} comparable to the luminosity in
the ZEUS analysis \cite{Chekanov:2007rh}. Second, H1 took data with low-$E_T^{jet}$ and 
high-$E_T^{jet}$ cuts, which allows for a comparison of
\cite{Aktas:2007hn} with \cite{Chekanov:2007rh}. The exact two kinematic ranges are given in \cite{h1dis08}.
%
%%%%%%%%%%%%%% Begin table 1 %%%%%%%%%%%%%%%%%%%%%%%%%%%%%%%%%%%%%%%%%%%
%\begin{table}
% \begin{tabular}{c|c}
% H1 low-$E_T^{jet}$ cuts & H1 high-$E_T^{jet}$ cuts\\
% \hline
% $Q^2<$ 0.01 GeV$^2$   & \\
% 0.3 $< y <$ 0.65     & \\
% $E_T^{jet1}>$ 5 GeV & $E_T^{jet1} >$ 7.5 GeV \\
% $E_T^{jet2}>$ 4 GeV & $E_T^{jet2} >$ 6.5 GeV \\
% $-1 < \eta^{jet1(2)} < 2$ &  $-1.5 < \eta^{jet1(2)} < 1.5$ \\
% $x_{\p} <$ 0.03      &  $x_{\p} <$ 0.025 \\
% $|t| <$ 1 GeV$^2$    & \\
% $M_Y <$ 1.6 GeV      & \\
% \end{tabular}
% \caption{Kinematic cuts applied in the most recent H1 analysis of diffractive
% dijet photoproduction \cite{29}. The high-$E_T^{jet}$ cuts are identical to
% the low-$E_T^{jet}$ cuts unless indicated otherwise.}
%\end{table}
%%%%%%%%%%%%%% End of table 1 %%%%%%%%%%%%%%%%%%%%%%%%%%%%%%%%%%%%%%%%%%
%
The ranges for the low-$E_T^{jet}$ cuts are as in the previous H1 analysis 
\cite{Aktas:2007hn} and for the high-$E_T^{jet}$ cuts are chosen as 
in the ZEUS analysis with two exceptions. In the ZEUS analysis the maximal cut 
on $Q^2$ is larger and the data are taken in an extended $y$-range. The 
definition of the various variables can be found in the H1 and ZEUS
publications
\cite{Aktas:2007hn,Chekanov:2007rh}. Very important is the cut on $x_{\p}$. It is kept small in both 
analyses in order for the pomeron exchange to be dominant. In the experimental 
analysis as well as in the
NLO calculations, jets are defined with the inclusive $k_T$-cluster algorithm
\cite{Ellis:1993tq,Catani:1993hr} in the laboratory frame. At least two jets are required with the 
respective cuts on $E_T^{jet1}$ and $E_T^{jet2}$, where $E_T^{jet1(2)}$ refers 
to the jet with the largest (second largest) $E_T^{jet}$. 
%As is well known, the lower
%limits on the jet $E_T$ are chosen asymmetric in order to avoid an infrared
%sensitivity in those NLO cross section computations, which are integrated
%over $E_T^{jet}$ \cite{31}.

Before we confront the calculated cross sections with the experimental
data, we correct them for hadronization effects. The hadronization
corrections are calculated by means of the LO RAPGAP Monte Carlo generator. 
The factors for the transformation of jets made up of stable hadrons 
to parton jets were supplied by the H1 collaboration \cite{h1dis08}. Our 
calculations are done with the `H1 2006 fit B' \cite{Aktas:2006hy} DPDFs, since
they give smaller diffractive dijet cross sections than with the `H1 2006 fit A'.
%The H1 collaboration constructed a third set of DPDFs, which is called the
%'H1 2007 fit jets'. This fit is obtained through a simultaneous fit to the
%diffractive inclusive and DIS dijet cross sections \cite{32}. In this fit 
%it is
%assumed that there is no factorization breaking in the diffractive DIS dijet 
%cross sections. Including these cross sections in the fit leads to additional 
%constraints, mostly for the diffractive gluon distribution. On average the
%'H1 2007 fit jets' is similar to the 'H1 2006 fit B' except for the gluon
%distribution at large momentum fraction and small factorization scale. In the
%following analysis we shall disregard this new DPDF set, since it would be
%compatible with the factorization test of the photoproduction data only, if we
%restricted these tests to the case that the resolved part has the breaking 
%and 
%not the direct part, which has the same theoretcal structute as the DIS 
%dijet cross sections. Results with the 'H1 2007 fit jets' can be found in
%\cite{29}. The `H1 2006 fits A,B' are based on $n_f=3$ massless 
%flavors. The production of charm quarks was treated in the Fixed-Flavor Number
%Scheme (FFNS) in NLO with non-zero charm quark. Instead of this extra 
%treatment
%of the charm contribution in the Pomeron we added a charm PDF in the Pomeron 
%as obtained in the `H1 2002 fit' \cite{9}, where the charm quark was also 
%considered to be massless. The bottom contribution was neglected. This 
%assumption simplifies the calculations considerably. Since the charm 
%contribution from the Pomeron is small, this should be a good approximation.
We then take $n_f=4$ with $\Lambda^{(4)}_{\overline{\rm MS}} = 0.347$ GeV, 
which corresponds to the value used in the DPDFs `H1 2006 fit A, B' \cite{Aktas:2006hy}. 
For the photon PDFs we have chosen the NLO GRV parameterization transformed 
to the $\overline{\rm MS}$ scheme \cite{Gluck:1991jc}.

%
%%%%%%%%%%%%%% Begin figure 1 %%%%%%%%%%%%%%%%%%%%%%%%%%%%%%%%%%%%%%%%%%
\begin{figure}
 \centering
 \includegraphics[width=0.325\columnwidth]{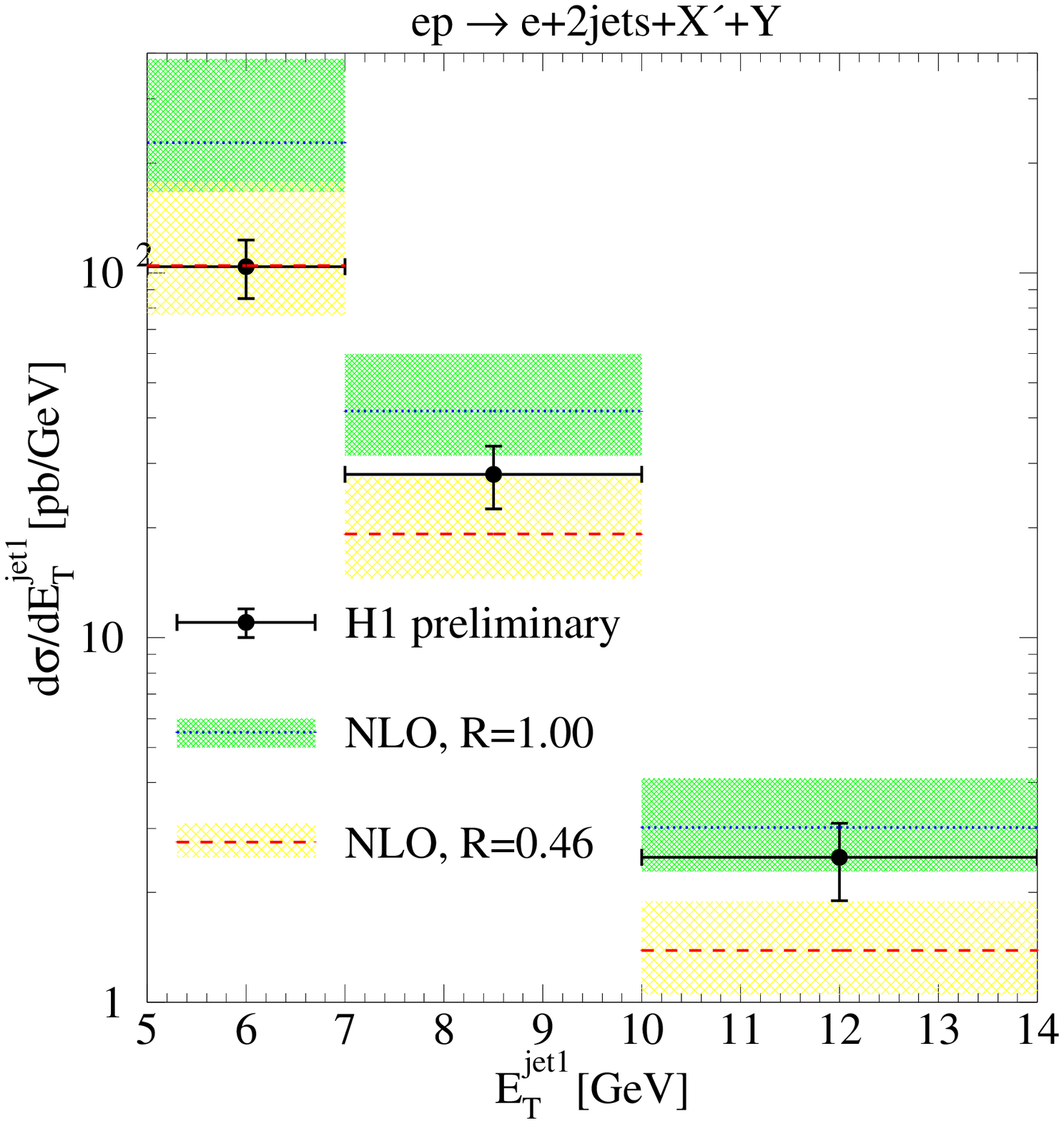}
 \includegraphics[width=0.325\columnwidth]{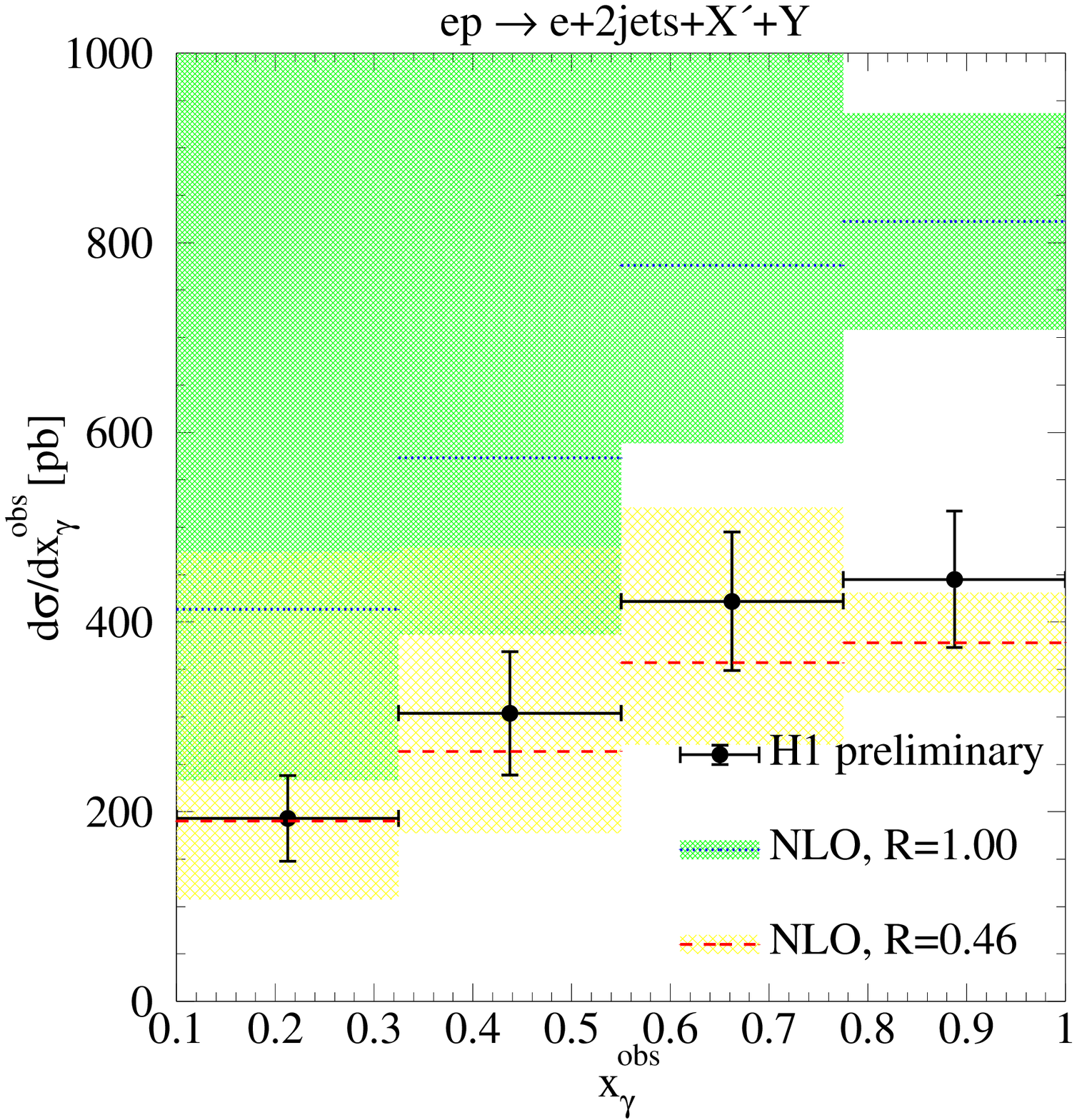}
 \includegraphics[width=0.325\columnwidth]{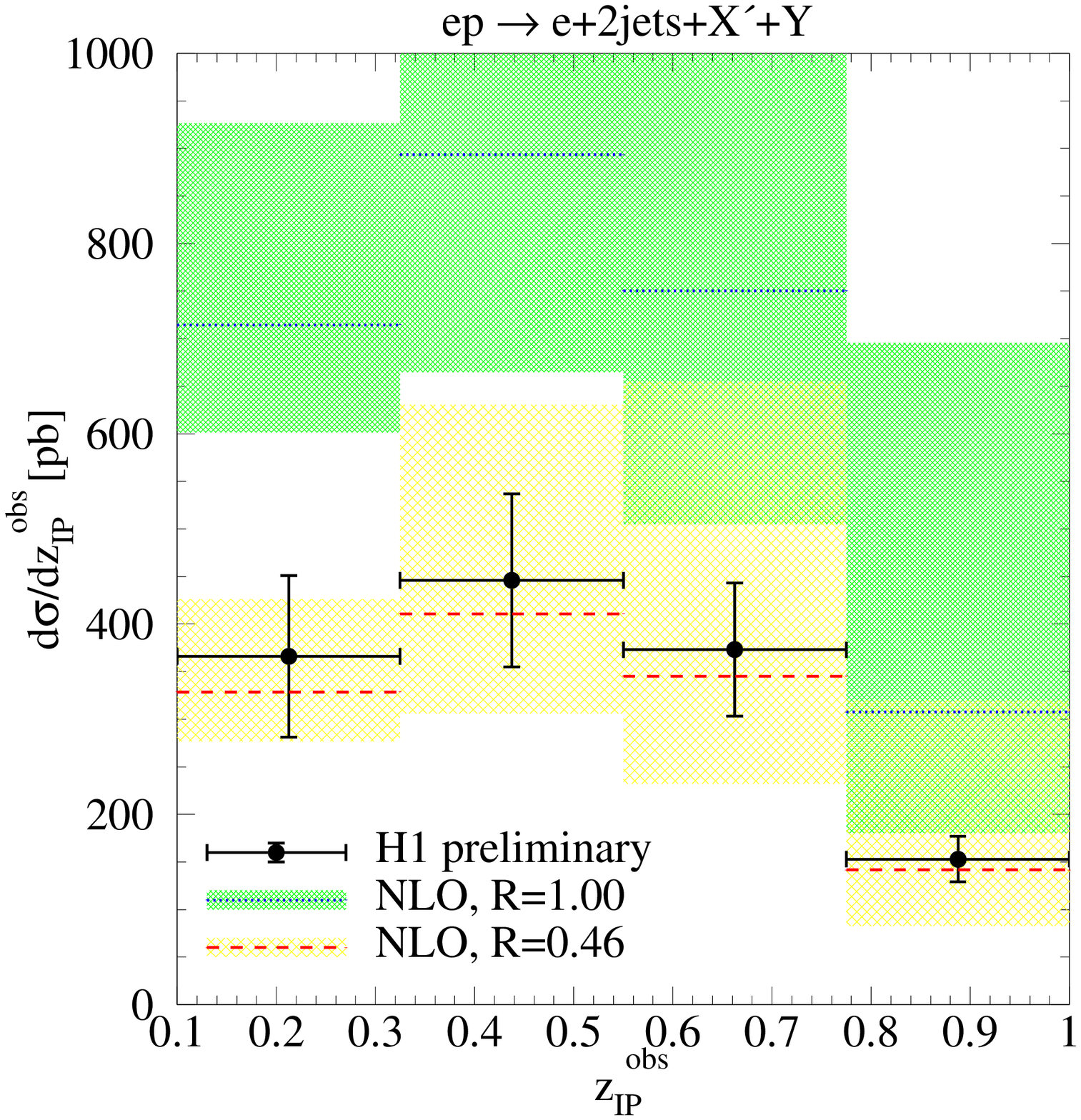}
 \caption{\label{fig:1}Differential cross sections for diffractive dijet
 photoproduction as measured by H1 with low-$E_T^{jet}$ cuts and compared to
 NLO QCD without ($R=1$) and with ($R=0.46$) global suppression 
 (color online).}
\end{figure}
%%%%%%%%%%%%%% End of figure 1 %%%%%%%%%%%%%%%%%%%%%%%%%%%%%%%%%%%%%%%%%
%
%
%%%%%%%%%%%%%% Begin figure 2 %%%%%%%%%%%%%%%%%%%%%%%%%%%%%%%%%%%%%%%%%%
\begin{figure}
 \centering
 \includegraphics[width=0.325\columnwidth]{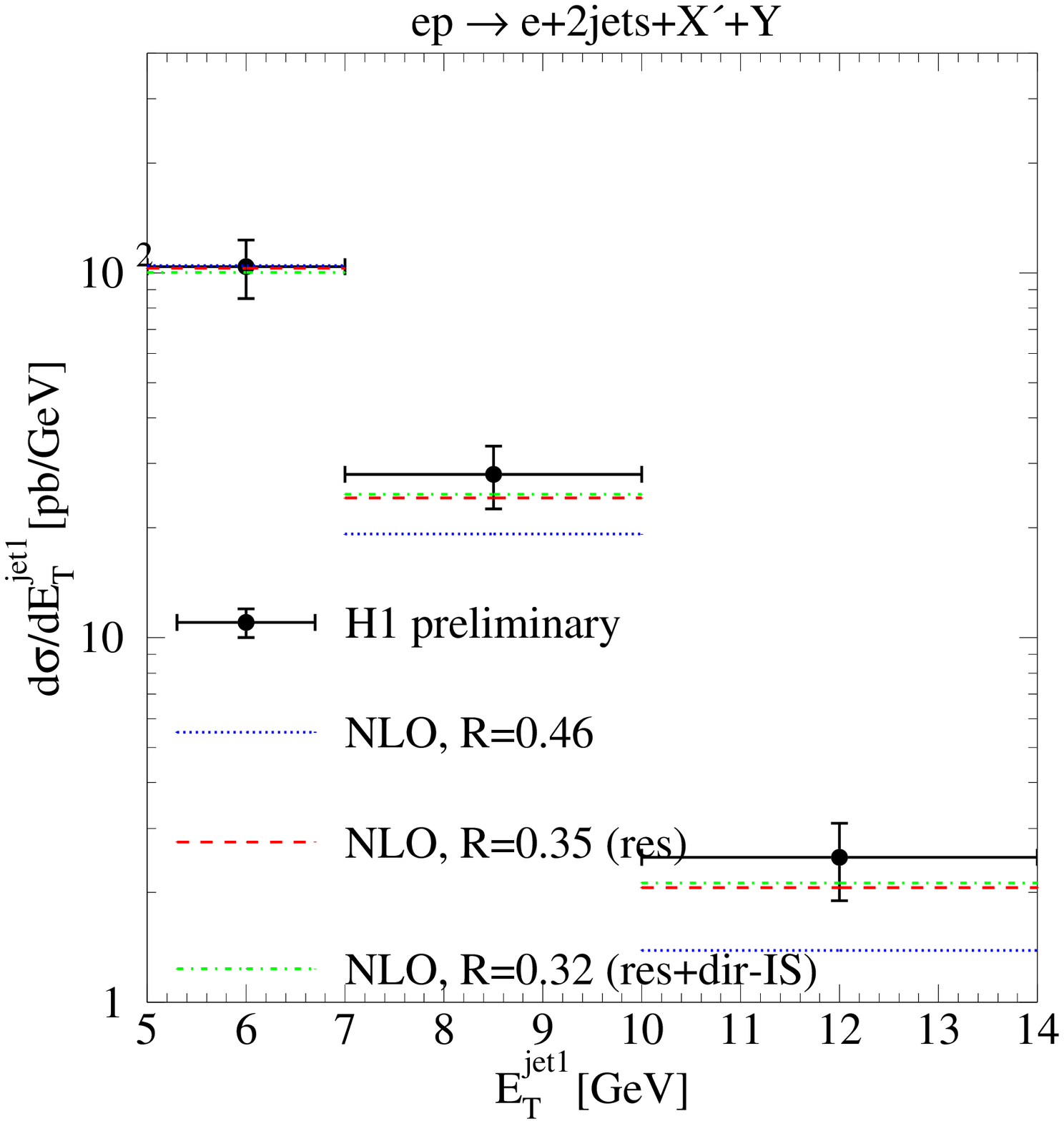}
 \includegraphics[width=0.325\columnwidth]{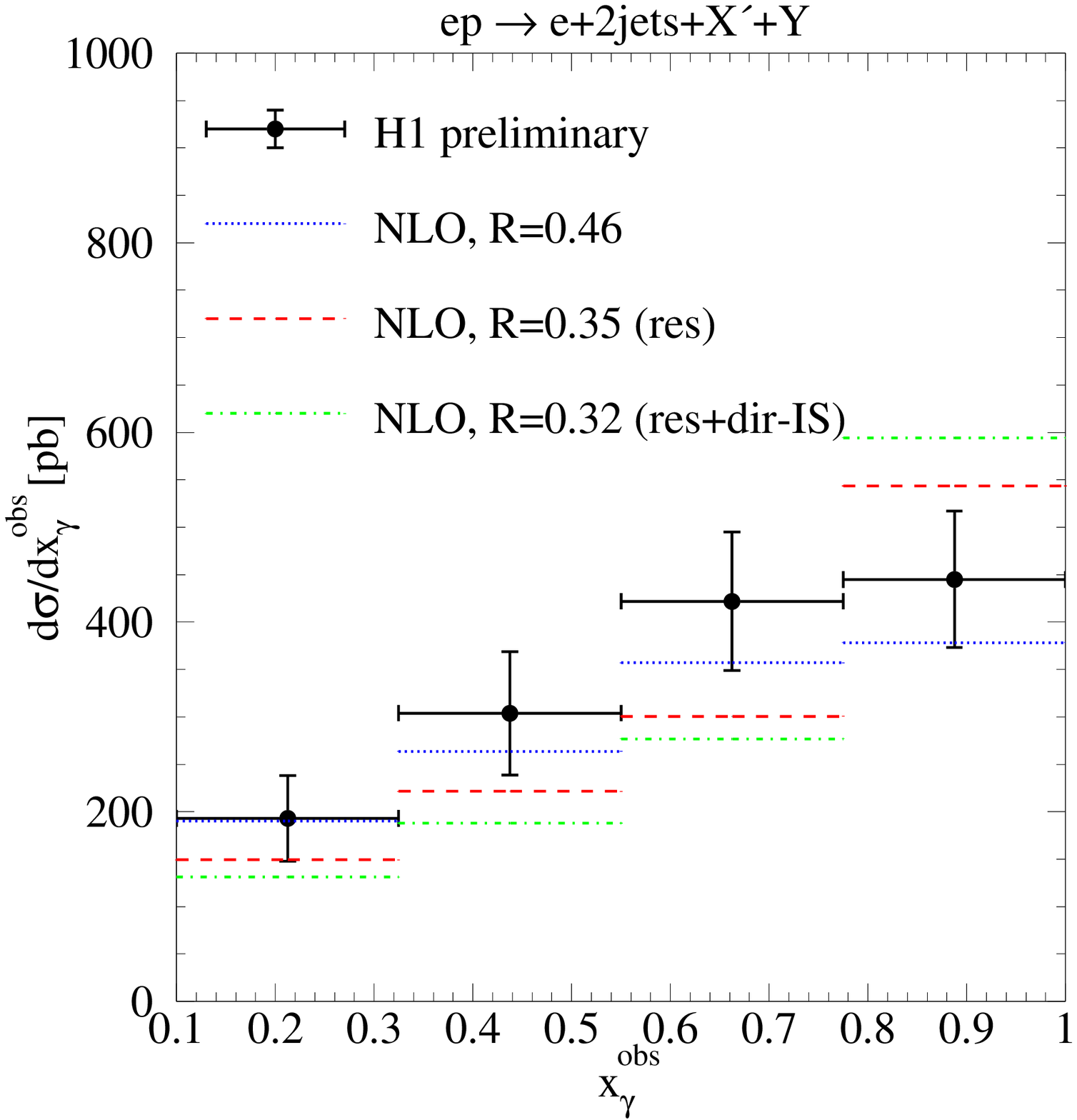}
 \includegraphics[width=0.325\columnwidth]{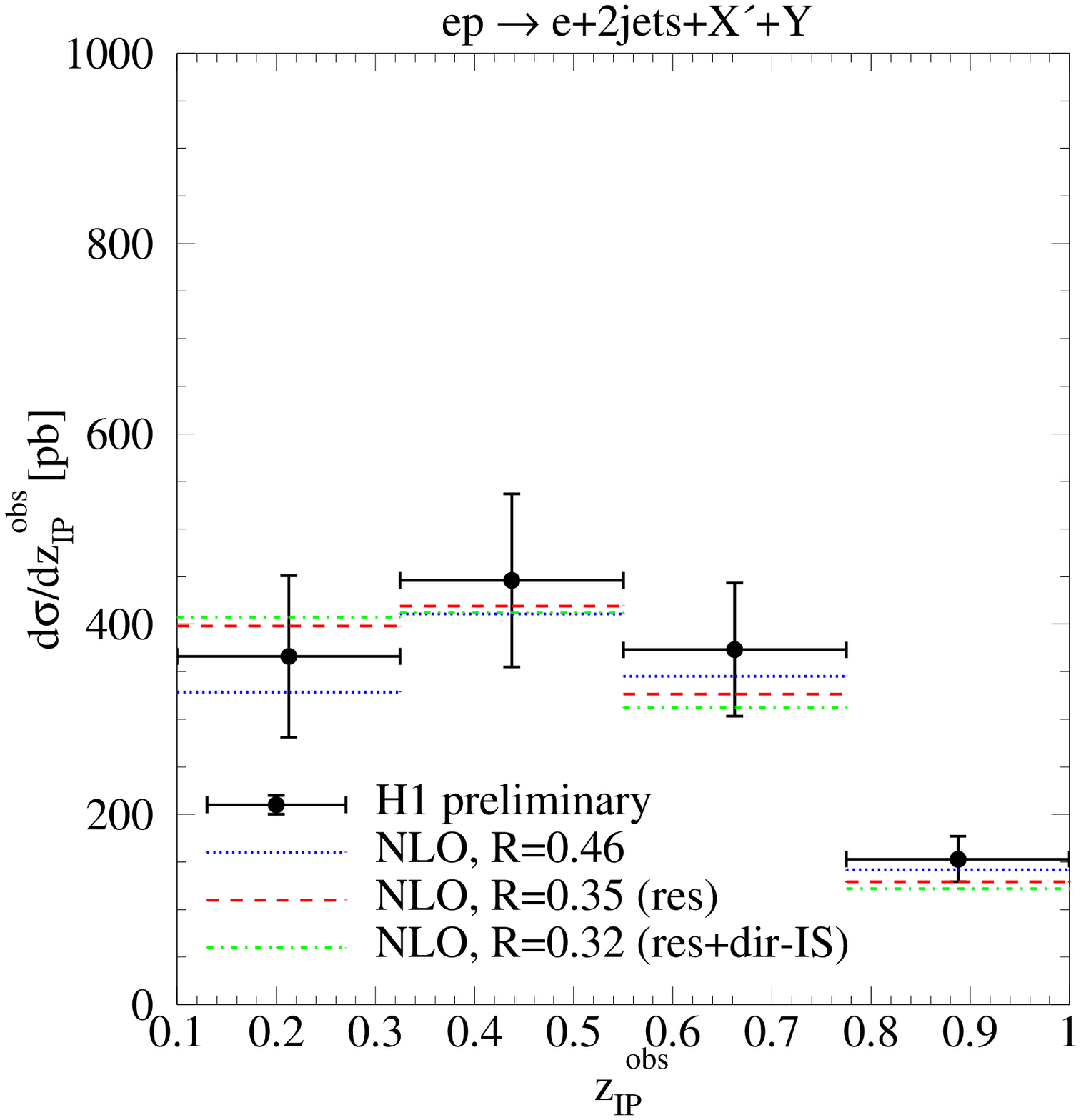}
 \caption{\label{fig:2}Differential cross sections for diffractive dijet
 photoproduction as measured by H1 with low-$E_T^{jet}$ cuts and compared to
 NLO QCD with global, resolved, and resolved/direct-IS suppression.}
\end{figure}
%%%%%%%%%%%%%% End of figure 2 %%%%%%%%%%%%%%%%%%%%%%%%%%%%%%%%%%%%%%%%%
%

As it is clear from the discussion of the various preliminary analyses of
the H1 and ZEUS collaborations, there are two questions which we would
like to answer from the comparison with the recent H1 and the ZEUS data.
The first question is whether a suppression factor, which differs substantially
from one, is needed to describe the data. The second question is whether the 
data are also consistent with a suppression factor applied to the resolved 
cross section only. For both suppression models it is also of interest whether
the resulting suppression factors are universal, i.e.\ whether they are 
independent of the kinematic variables of the process. To give an answer to 
these two questions we calculated first the cross 
sections with no suppression factor ($R = 1$ in the following figures) with a
theoretical error obtained from varying the common scale of renormalization and
factorization by factors of 0.5  and 2 around the default value (highest
$E_T^{jet}$). In a second step we show the results  
for the same differential cross sections with a global suppression factor,
adjusted to $d\sigma/dE_T^{jet1}$ at the smallest $E_T^{jet1}$-bin. As in the
experimental analysis \cite{h1dis08}, we consider the differential cross sections
in  the variables $x_{\gamma}^{obs}$, $z_{\p}^{obs}$, $\log_{10}(x_{\p})$,
$E_T^{jet1}$,  $M_X$, $M_{12}$, $\overline{\eta}^{jets}$, $|\Delta
\eta^{jets}|$ and $W$ \cite{kk08tbp}. 
%The definition of the variables is given in the
%experimental papers \cite{27,28,29} or in our earlier work \cite{17,18,34}. 
%In 
%the latter references also the relevant formulas for the calculation of the 
%dijet cross sections can be found.
Here we show only a selection, i.e.\ the cross sections as a function of 
$E_T^{jet1}$, $x_{\gamma}^{obs}$ and $z_{\p}^{obs}$.
For the low-$E_T^{jet}$ cuts, the resulting suppression factor is 
$R = 0.46\pm 0.14$, which gives in the lowest $E_T^{jet1}$-bin a cross section 
equal to the experimental data point. The error comes from the combined 
experimental statistical and systematic error. The theoretical error due to the
scale variation is taken into account when comparing to the three 
distributions. The results of this comparison are shown in Figs.\ 1a-c.
With the exception of Fig.\ 1a , where the comparison of 
$d\sigma/dE_T^{jet1}$ is shown, the other two plots are 
such that the data points lie outside the error
band based on the scale variation for the unsuppressed case. However,
the predictions with suppression $R = 0.46$ agree nicely with the
data inside the error bands from the scale variation. Most of the data
points even agree with the $R = 0.46$ predictions inside the much smaller
experimental errors. 
%The $M_X$ distributions agrees only for the second bin.
%The reason for the disagreeent of the two othe $M_X$ bins might be that in the
%theoretical results the variable $M_X$ is defined with out the remnant
%contributions, which, however, are taken into account in the experimental
%definition of $M_X$. An exceptions is the cross section 
%$d\sigma/dE_T^{jet1}$. 
In $d\sigma/dE_T^{jet1}$ (see Fig.\ 1a) the predictions for the second and 
third bins lie outside the data points with their errors. For $R = 1$ and 
$R=0.46$ this cross sections falls off stronger with increasing $E_T^{jet1}$  
than the data, the normalization being of course about two times larger for 
$R=1$. In particular, the third data point agrees with the $R = 1$ prediction.
This means that the suppression decreases with increasing $E_T^{jet1}$
(see also Fig.\ \ref{fig:11} below). This
behavior was already apparent when we analyzed the first preliminary H1 data
\cite{Klasen:2004tza,Klasen:2004qr}. Such a behavior points in the direction that a suppression of 
the resolved cross section only would give better agreement with the data, as 
we shall see below. 
%The same  observations can be made by looking at 
%$d\sigma/dM_{12}$ in Fig.\ 3f. 
The survival probability $R = 0.46 \pm 0.14$ agrees with the result in 
\cite{h1dis08}, which quotes $R=0.51 \pm 0.01$ (stat.) $\pm$ 0.10 (syst.),
determined by fitting the integrated cross section. From our comparison we 
conclude that the low-$E_T^{jet}$  data show a global suppression of the order
of  two in complete agreement with the results \cite{Klasen:2004tza,Klasen:2004qr} and \cite{Aktas:2007hn} 
based on earlier preliminary and final H1 data \cite{Aktas:2007hn}. 

%
%%%%%%%%%%%%%% Begin figure 3 %%%%%%%%%%%%%%%%%%%%%%%%%%%%%%%%%%%%%%%%%%
\begin{figure}
 \centering
 \includegraphics[width=0.325\columnwidth]{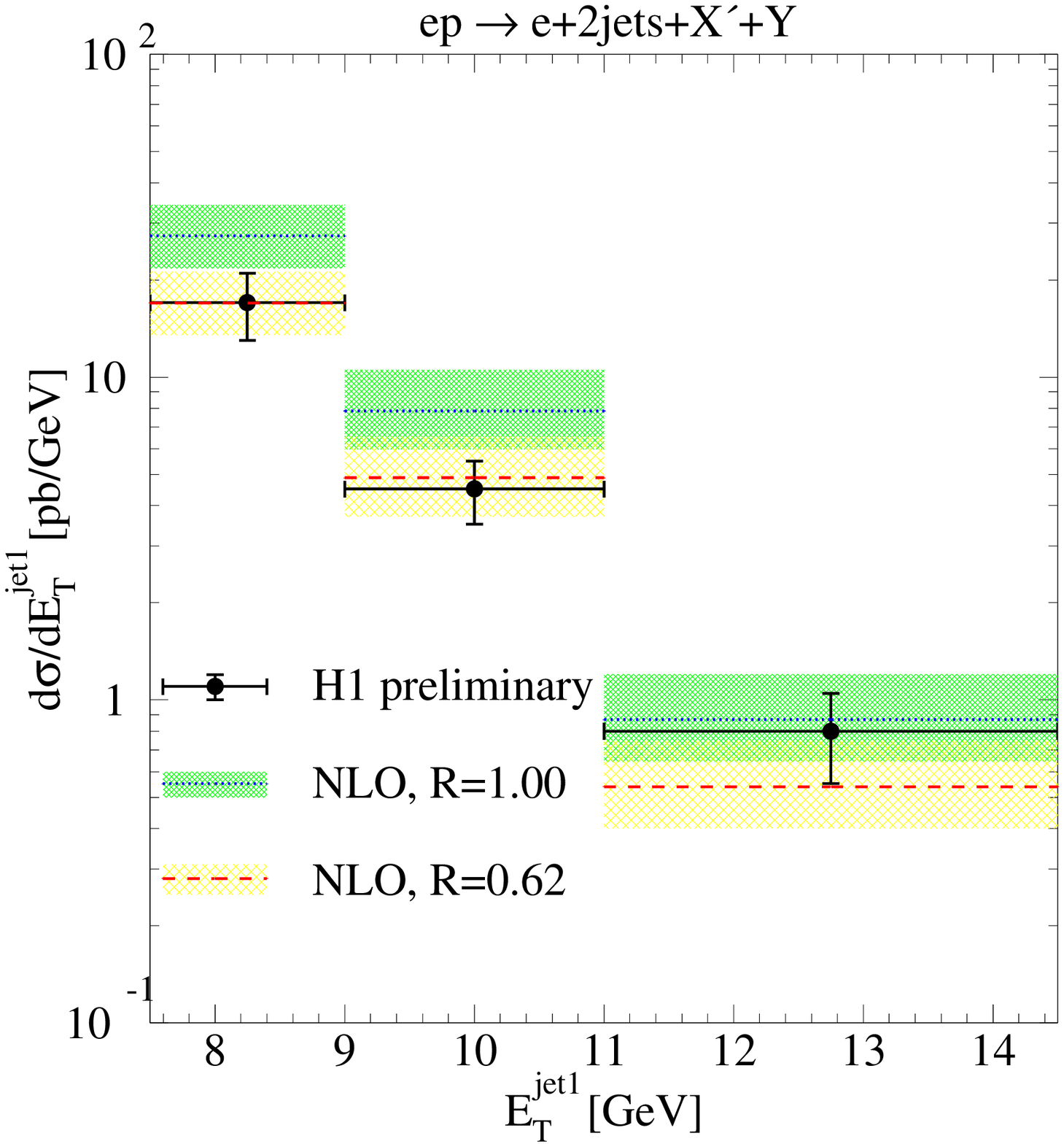}
 \includegraphics[width=0.325\columnwidth]{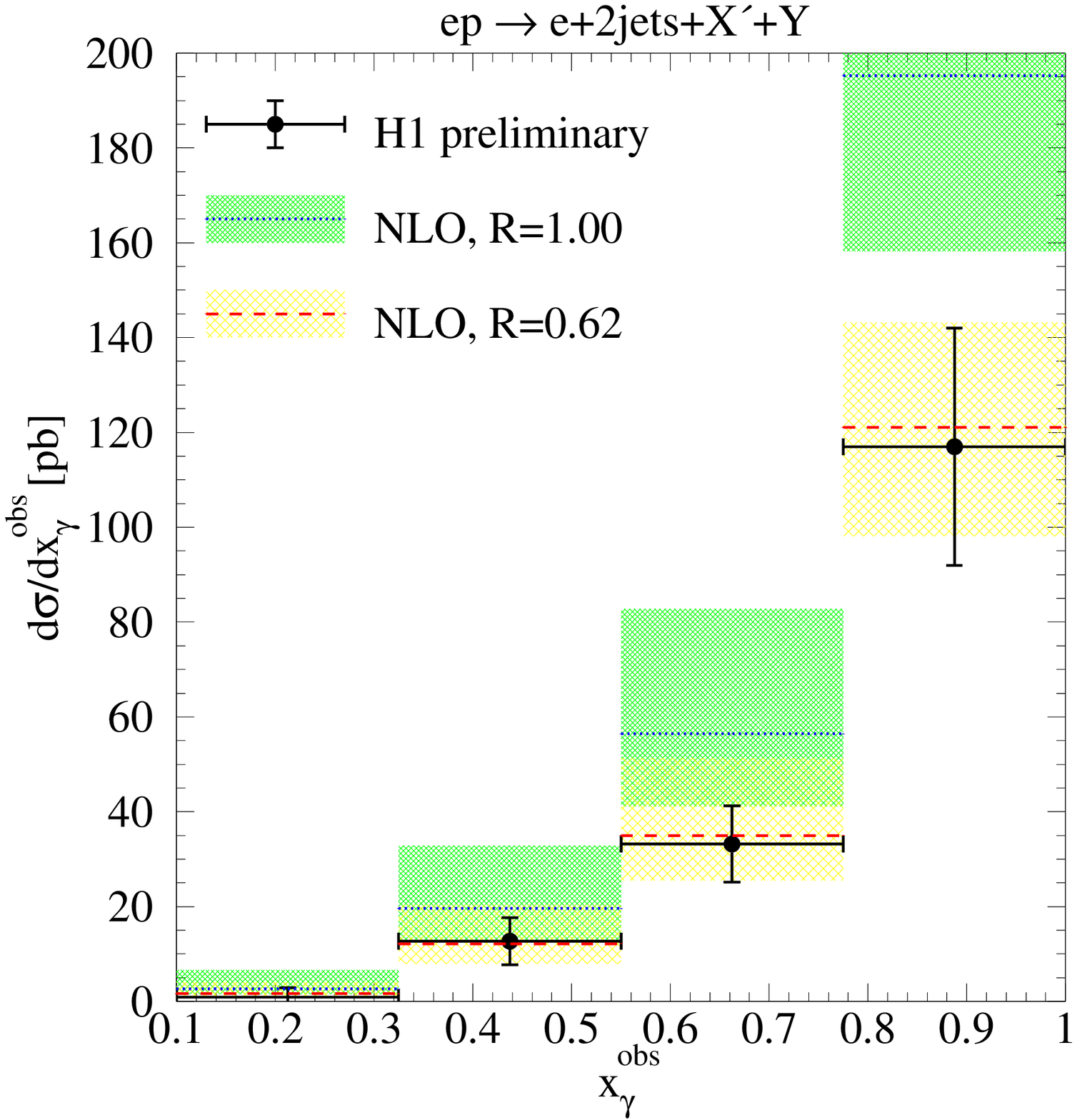}
 \includegraphics[width=0.325\columnwidth]{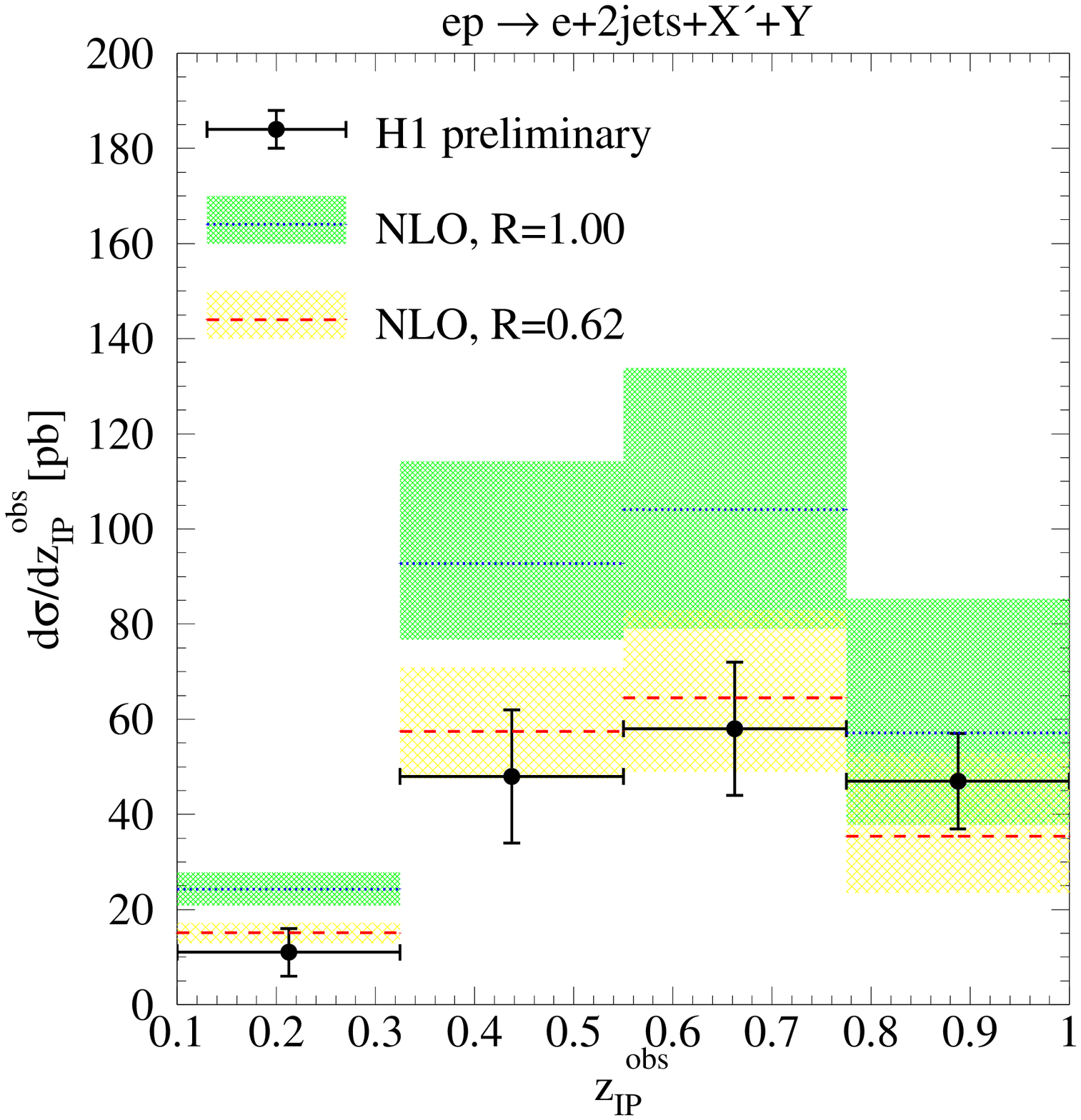}
 %\includegraphics[width=0.325\columnwidth]{fig5d}
 %\includegraphics[width=0.325\columnwidth]{fig5e}
 %$\includegraphics[width=0.325\columnwidth]{fig5f}
 %\includegraphics[width=0.325\columnwidth]{fig5g}
 %\includegraphics[width=0.325\columnwidth]{fig5h}
 %\includegraphics[width=0.325\columnwidth]{fig5i}
 \caption{\label{fig:3}Differential cross sections for diffractive dijet
 photoproduction as measured by H1 with high-$E_T^{jet}$ cuts and compared to 
 NLO QCD without ($R=1$) and with ($R=0.62$) global suppression 
 (color online).}
\end{figure}
%%%%%%%%%%%%%% End of figure 3 %%%%%%%%%%%%%%%%%%%%%%%%%%%%%%%%%%%%%%%%%
%
%
%%%%%%%%%%%%%% Begin figure 4 %%%%%%%%%%%%%%%%%%%%%%%%%%%%%%%%%%%%%%%%%%
\begin{figure}
 \centering
 \includegraphics[width=0.325\columnwidth]{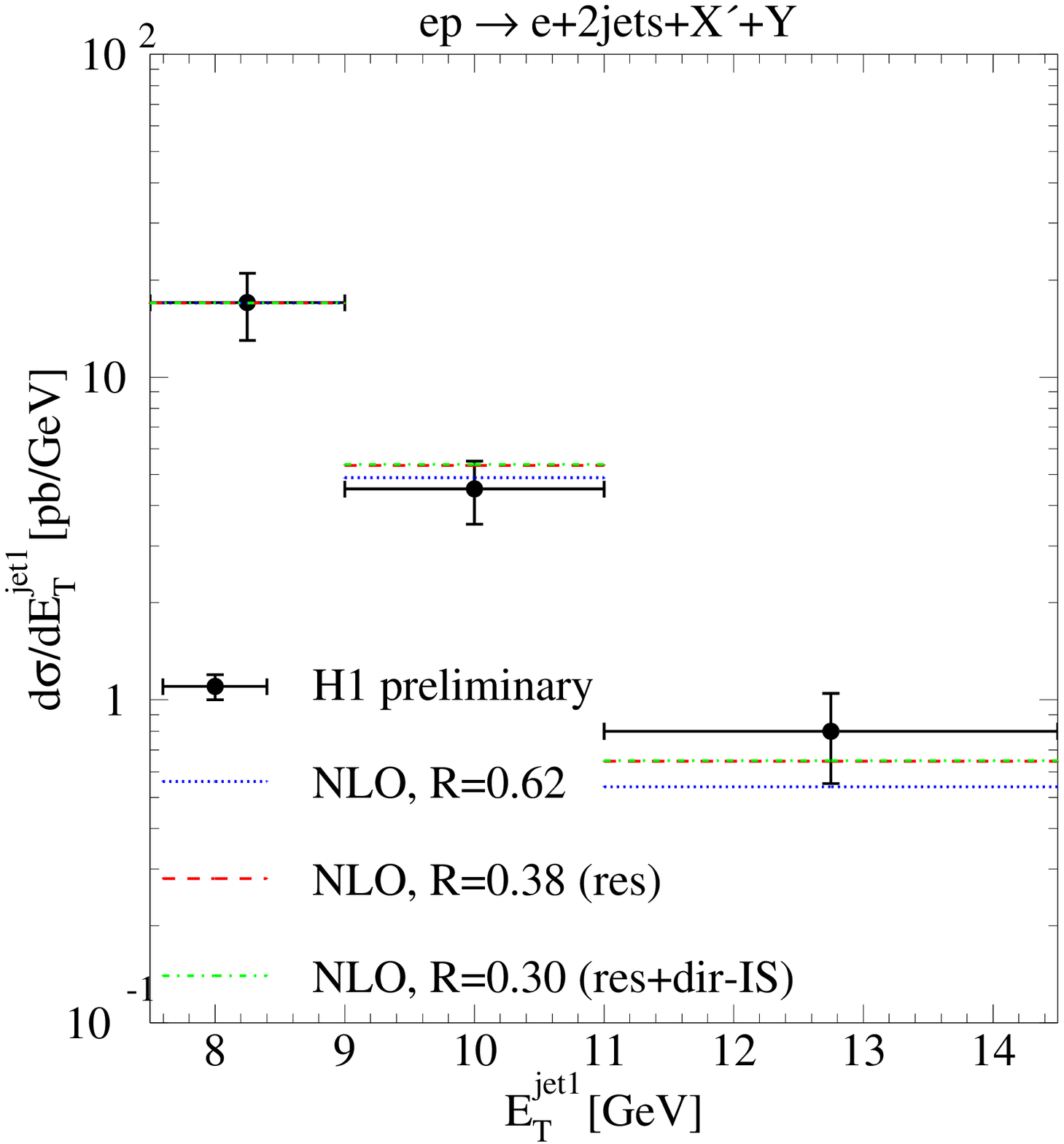}
 \includegraphics[width=0.325\columnwidth]{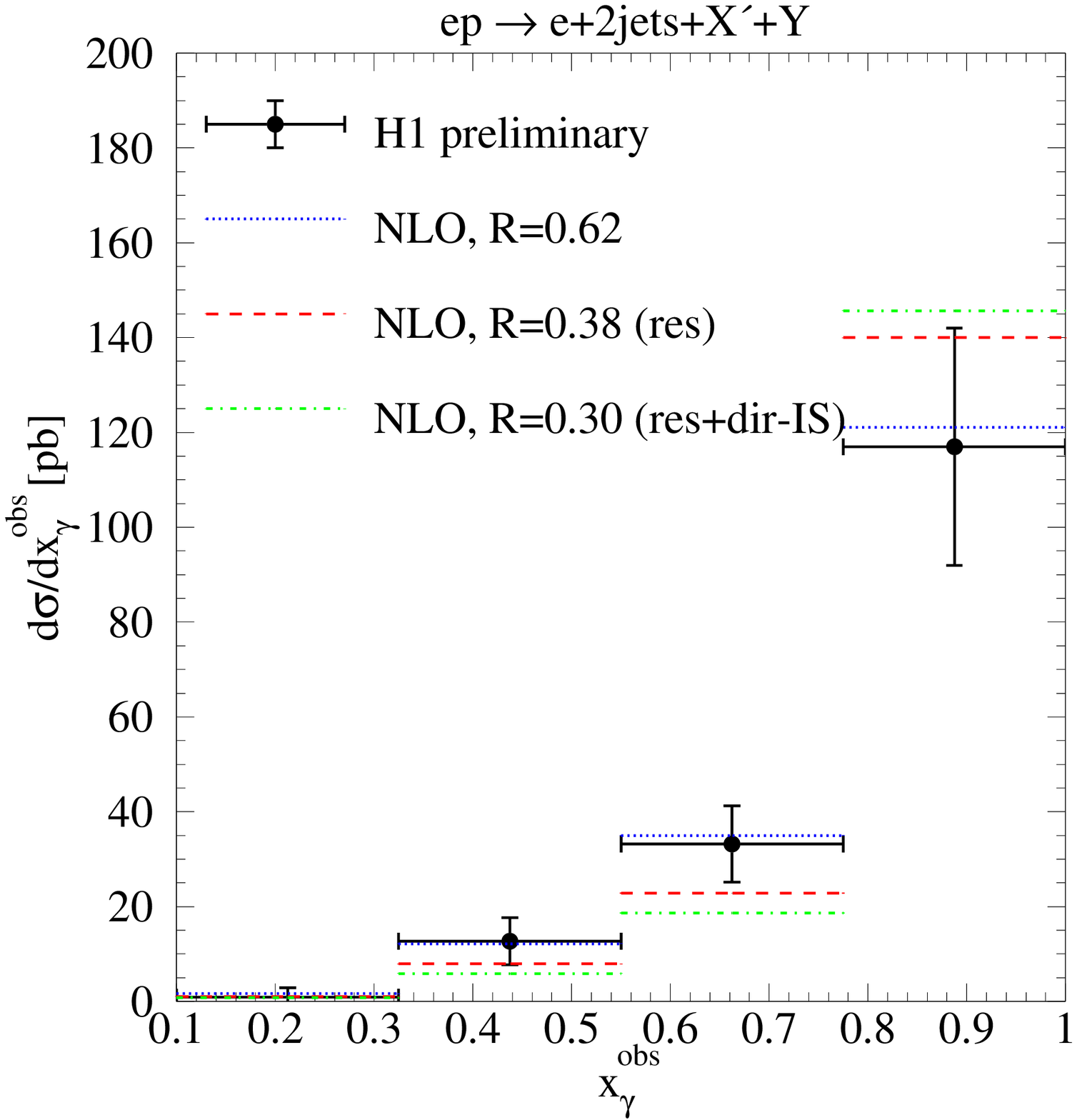}
 \includegraphics[width=0.325\columnwidth]{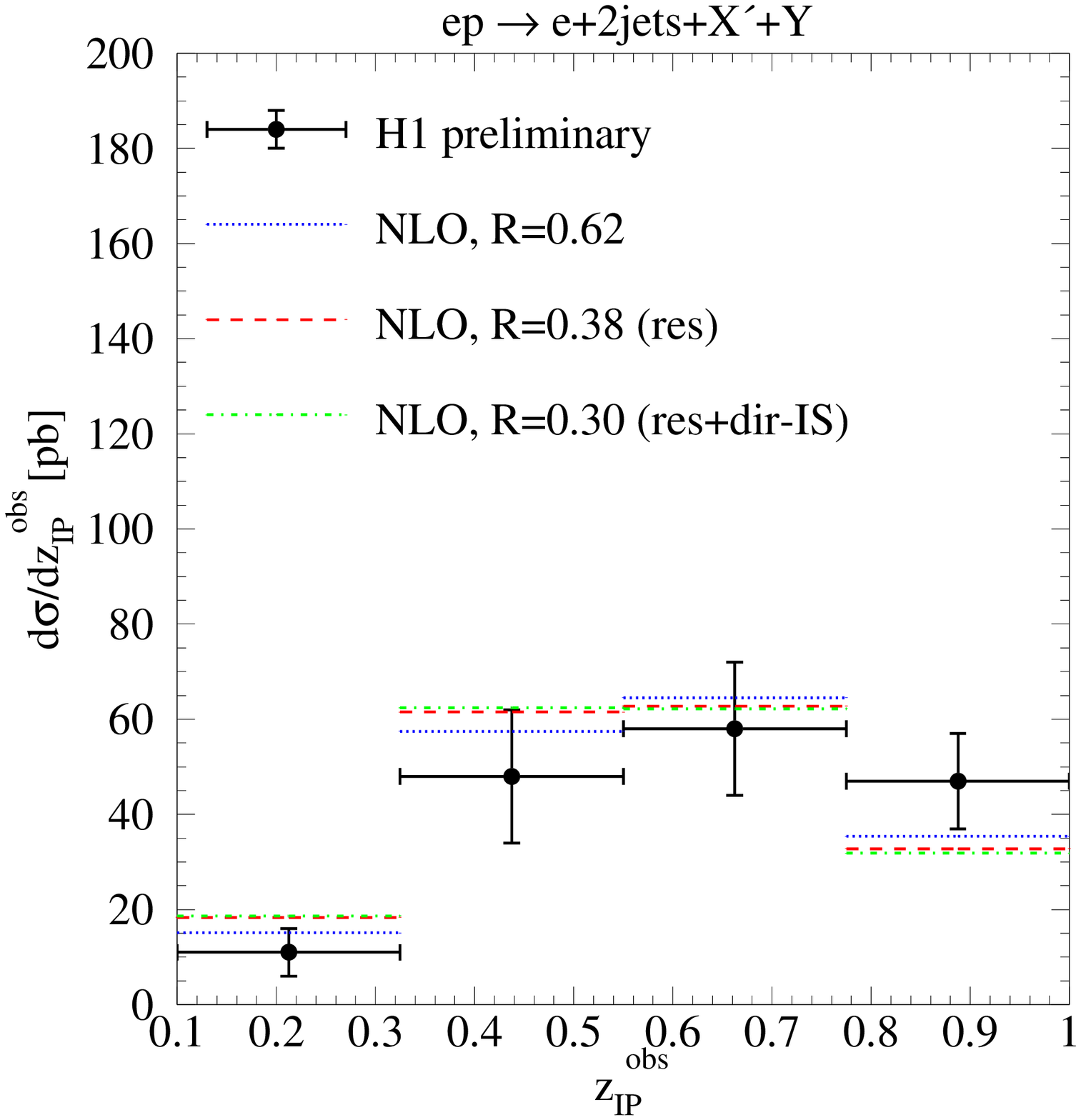}
 \caption{\label{fig:4}Differential cross sections for diffractive dijet
 photoproduction as measured by H1 with high-$E_T^{jet}$ cuts and compared to
 NLO QCD with global, resolved, and resolved/direct-IS suppression.}
\end{figure}
%%%%%%%%%%%%%% End of figure 4 %%%%%%%%%%%%%%%%%%%%%%%%%%%%%%%%%%%%%%%%%
%

Next we want to answer the second question, whether the data could be consistent
with a suppression of the resolved component only. For this purpose we have
calculated the cross sections in two additional versions: (i) suppression of the
resolved cross section and (ii) suppression of the resolved cross section plus the
NLO direct part which depends on the factorization scale at the photon vertex
\cite{Klasen:2005dq}. The suppression factors needed for the two versions will,
of course, be different. We determine them again by fitting the measured
$d\sigma/dE_T^{jet1}$ for the lowest $E_T^{jet1}$-bin (see Fig.\ 2a). Then, the
suppression factor for version (i) is $R = 0.35$ (denoted res in the figures), and
for version (ii) it is $R = 0.32$ (denoted res+dir-IS). The results for
$d\sigma/dE_T^{jet1}$, $ d\sigma/dx_{\gamma}^{obs}$ and $d\sigma/dz_{\p}^{obs}$
are shown in Figs.\ 2a-c, while the six other distributions can be found in
\cite{kk08tbp}.
We also show the global (direct and resolved) suppression prediction with
$R = 0.46$ already shown in Figs.\ 1a-c. For the cross section as a function of
$z_{\p}^{obs}$, the agreement with the global suppression ($R = 0.46$) and the
resolved suppression ($R = 0.35$ or $R = 0.32$) is comparable. For
$d\sigma/dE_T^{jet1}$, the agreement improves considerably for the resolved
suppression only (note the logarithmic scale in Fig.\ 2a). The global suppression
factor could, of course, be $E_T$-dependent, although we see no theoretical reason
for such a dependence. For $d\sigma/dx_{\gamma}^{obs}$, which is usually
considered as the characteristic
distribution for distinguishing global versus resolved suppression, the agreement
with resolved suppression does not improve. Unfortunately, this cross section has
the largest hadronic corrections of the order of $(25-30)\%$ \cite{h1dis08}.
Second, also for the usual photoproduction of dijets the comparison between data
and theoretical results has similar problems in the large $x_{\gamma}^{obs}$-bin
\cite{Aktas:2006qe}, although the $E_T^{jet}$-cut is much larger there. 
 %We also checked for two distributions whether the predictions for resolved 
%suppression depend on the chosen diffractive PDFs. For this purpose we have 
%calculated for the two cases $d\sigma/dz_{\p}^{obs}$ and $d\sigma/dE_T^{jet1}$ 
%the cross sections with the `H1 2006 fit A' parton distributions
%\cite{4}. The results are compared in Figs.\ 5a and b to the results with
%
%%%%%%%%%%%%%% Begin figure 5 %%%%%%%%%%%%%%%%%%%%%%%%%%%%%%%%%%%%%%%%%%
%\begin{figure}
% \centering
% \includegraphics[width=0.495\columnwidth]{fig4a}
% \includegraphics[width=0.495\columnwidth]{fig4b}
% \caption{\label{fig:5}Differential cross sections for diffractive dijet
% photoproduction as measured by H1 with low-$E_T^{jet}$ cuts and compared to
% NLO QCD with resolved suppression and two different DPDFs.}
%\end{figure}
%%%%%%%%%%%%%% End of figure 5 %%%%%%%%%%%%%%%%%%%%%%%%%%%%%%%%%%%%%%%%%
%
%the `H1 2006 fit B' and the experimental data. Of course, since the `H1 2006
%fit A' PDF's have a larger gluon component at large $z$, the cross sections 
%are
%$larger and therefore need a larger suppression factor $R = 0.26$. From Figs.\
%5a and b we conclude that there is no appreciable dependence on the chosen
%DPDFs. 
In total, we are tempted to conclude from the comparisons in  Figs.\
2a-c that the predictions with a resolved-only (or
resolved+direct-IS) suppression are consistent with the new low-$E_T^{jet}$
H1 data \cite{h1dis08} and the survival probability is $R=0.35$ (only resolved
suppression) and $R=0.32$ (resolved plus direct-IS suppression), respectively.

%
%%%%%%%%%%%%%% Begin figure 11 %%%%%%%%%%%%%%%%%%%%%%%%%%%%%%%%%%%%%%%%%
\begin{figure}
 \centering
 \includegraphics[width=0.325\columnwidth]{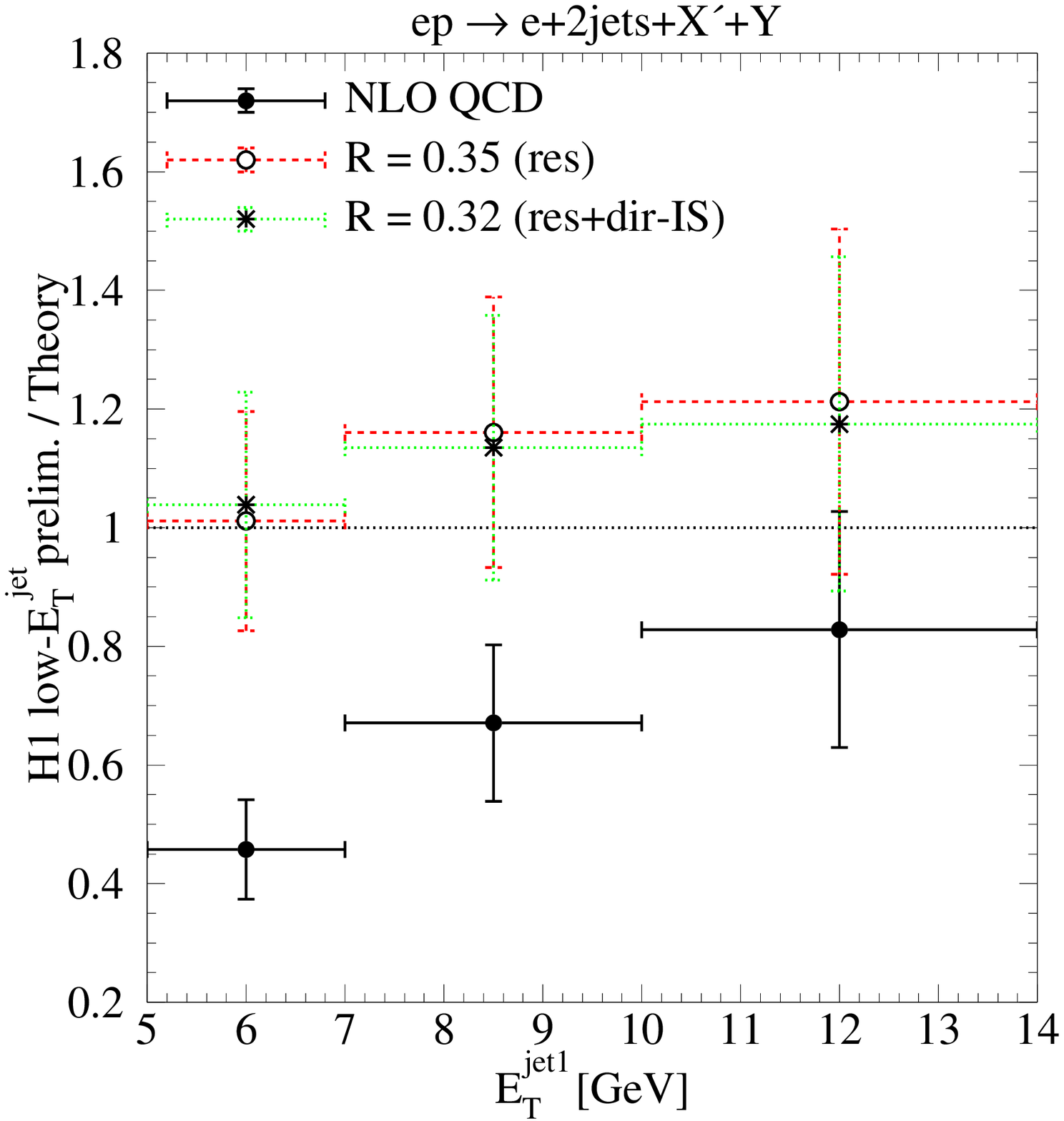}
 \includegraphics[width=0.325\columnwidth]{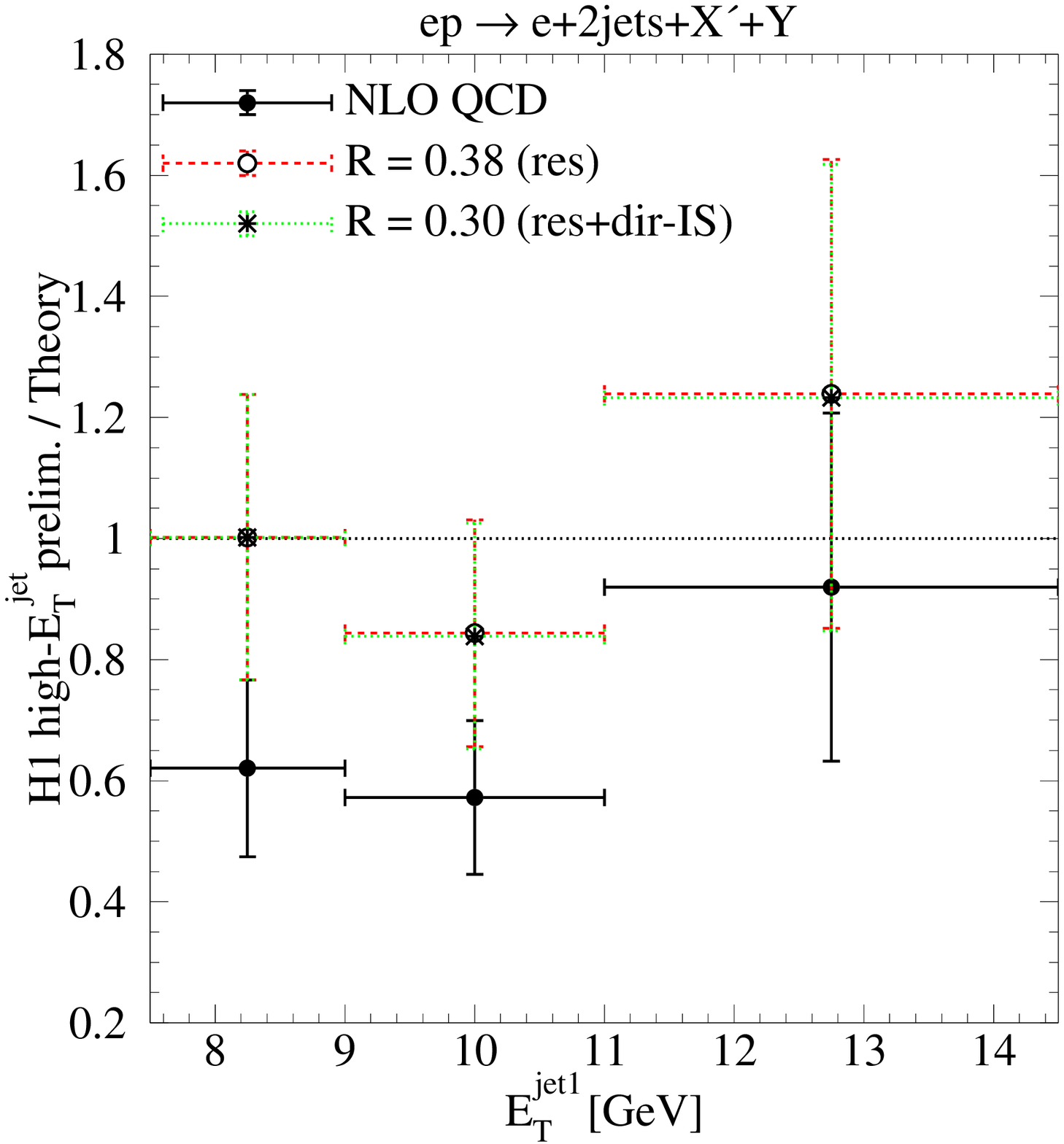}
 \caption{\label{fig:11}Ratio of the $E_T^{jet1}$-distribution as measured
 by H1 with low-$E_T^{jet}$ (left) and high-$E_T^{jet}$ cuts (right) to the
 NLO QCD prediction without (full), with resolved-only (dashed), and with
 additional direct IS suppression (dotted).}
\end{figure}
%%%%%%%%%%%%%% End of figure 11 %%%%%%%%%%%%%%%%%%%%%%%%%%%%%%%%%%%%%%%%
%

The same comparison of the high-$E_T^{jet}$ data of H1 \cite{h1dis08} with the 
various theoretical predictions is shown in the following figures. The global
suppression factor is obtained again from a fit to the smallest
$E_T^{jet1}$-bin.
It is equal to $R = 0.62 \pm 0.16$, again in agreement with the H1 result
$R=0.62\pm0.03$ (stat.) $\pm$ 0.14 (syst.) \cite{h1dis08}. 
The same cross sections as for the low-$E_T^{jet}$ comparison are shown in Figs.\
3a-c for the two cases $R = 1$ (no suppression) and $R = 0.62$
(global suppression), while the six others can again be found in
\cite{kk08tbp}. As before with the exception of $d\sigma/dE_T^{jet1}$ 
and $d\sigma/dM_{12}$ (not shown), most of the data points lie 
outside the $R = 1$ results with their
error bands and agree with the suppressed prediction with $R = 0.62$ inside
the respective errors. However, compared to the results in Figs.\ 1a-c
the distinction between the $R = 1$ band and the $R = 0.62$ band
and the data is somewhat less pronounced, which is due to the larger 
suppression factor. We also tested the prediction for the
resolved (resolved+direct-IS) suppression, which is shown in Figs.\ 4a-c. The 
suppression factor fitted to the smallest  bin
came out as $R = 0.38$ (res) and $R = 0.30$ (res+dir-IS), which are almost 
equal to the corresponding suppression factors derived from the low-$E_T^{jet}$
data. In most of the comparisons it is hard to observe any preference for the 
global against the pure resolved (resolved plus direct-IS) suppression. We
remark that the suppression factor for the global suppression is increased
by $35\%$, if we go from the low-$E_T^{jet}$ to the high-$E_T^{jet}$ data,
whereas for the resolved suppression this increase is only $9\%$. Under the
assumption that the suppression factor should not depend on
$E_T^{jet1}$, we would conclude that the resolved suppression
would be preferred, as can also be seen from Fig.\ \ref{fig:11}.
%In Figs.\ 8a and b we tested the resolved
%
%%%%%%%%%%%%%% Begin figure 8 %%%%%%%%%%%%%%%%%%%%%%%%%%%%%%%%%%%%%%%%%%
%\begin{figure}
% \centering
% \includegraphics[width=0.495\columnwidth]{fig7a}
% \includegraphics[width=0.495\columnwidth]{fig7b}
% \caption{\label{fig:8}Differential cross sections for diffractive dijet
% photoproduction as measured by H1 with high-$E_T^{jet}$ cuts and compared to
% NLO QCD with resolved suppression and two different DPDFs.}
%\end{figure}
%%%%%%%%%%%%%% End of figure 8 %%%%%%%%%%%%%%%%%%%%%%%%%%%%%%%%%%%%%%%%%
%
%suppression model against the choice of the two DPDFs, fit A versus fit B,
%with the result that this dependence is weak if we adjust the suppression
%pinefactor, which is $R = 0.16$ for the `H1 2006 fit A'. The general
%
% conclusions
%from the high-$E_T^{jet}$ comparison are very much the same as from the
%low-$E_T^{jet}$ comparison. 
A global suppression is definitely observed also in the high-$E_T^{jet}$ data
and the version with resolved suppression explains the data almost as well as 
with the global suppression.

In Fig.\ \ref{fig:11} we show the ratio of of the $E_T^{jet1}$-distribution as
measured by H1 to the NLO QCD prediction without (full), with resolved-only
(dashed), and with additional direct IS suppression (dotted). Within the
experimental errors, obviously only the former, but not the latter are
$E_T^{jet}$-dependent.

\section{Comparison with ZEUS data}
%
%%%%%%%%%%%%%% Begin figure 9 %%%%%%%%%%%%%%%%%%%%%%%%%%%%%%%%%%%%%%%%%%
\begin{figure}
 \centering
 \includegraphics[width=0.325\columnwidth]{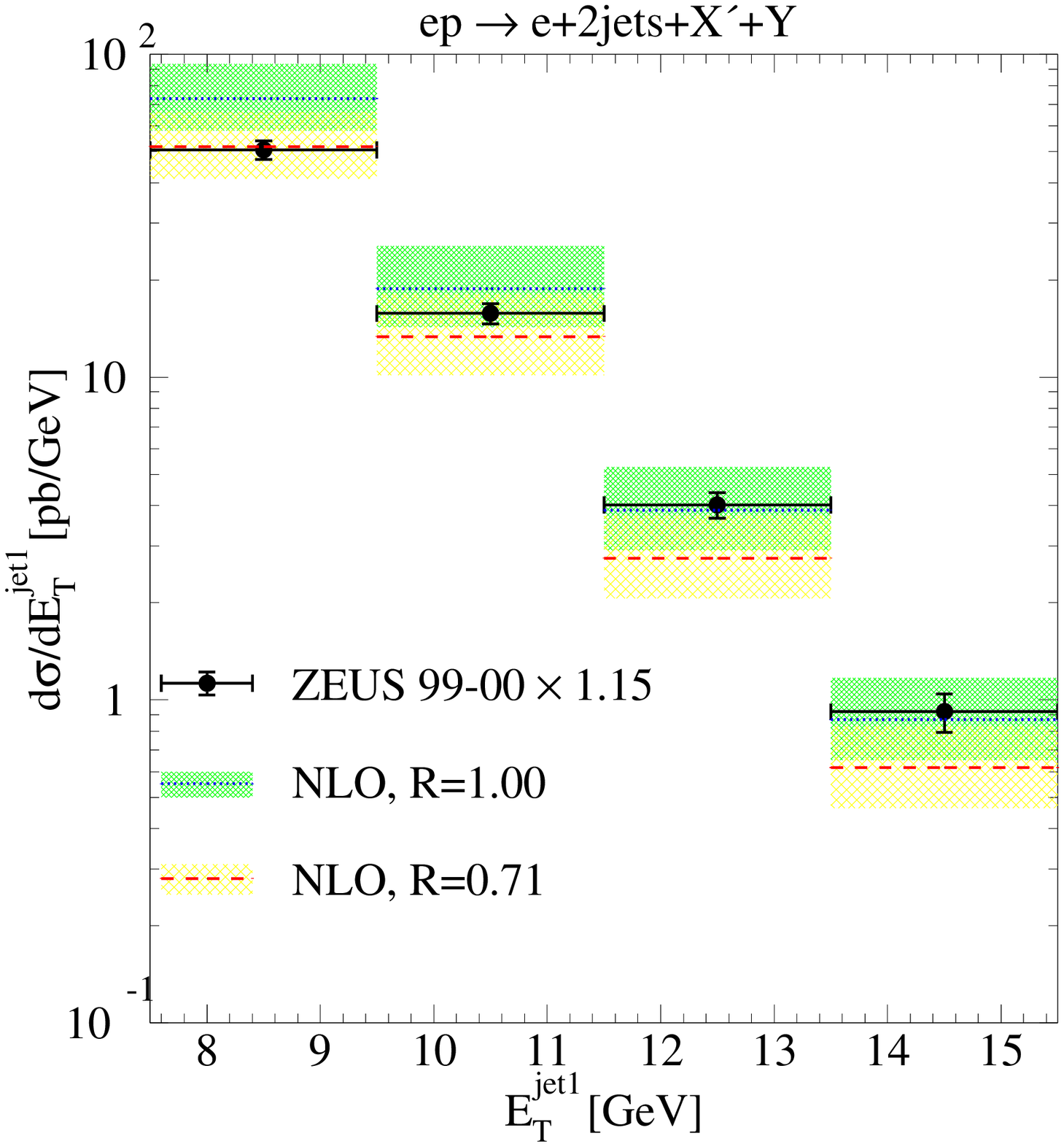}
 \includegraphics[width=0.325\columnwidth]{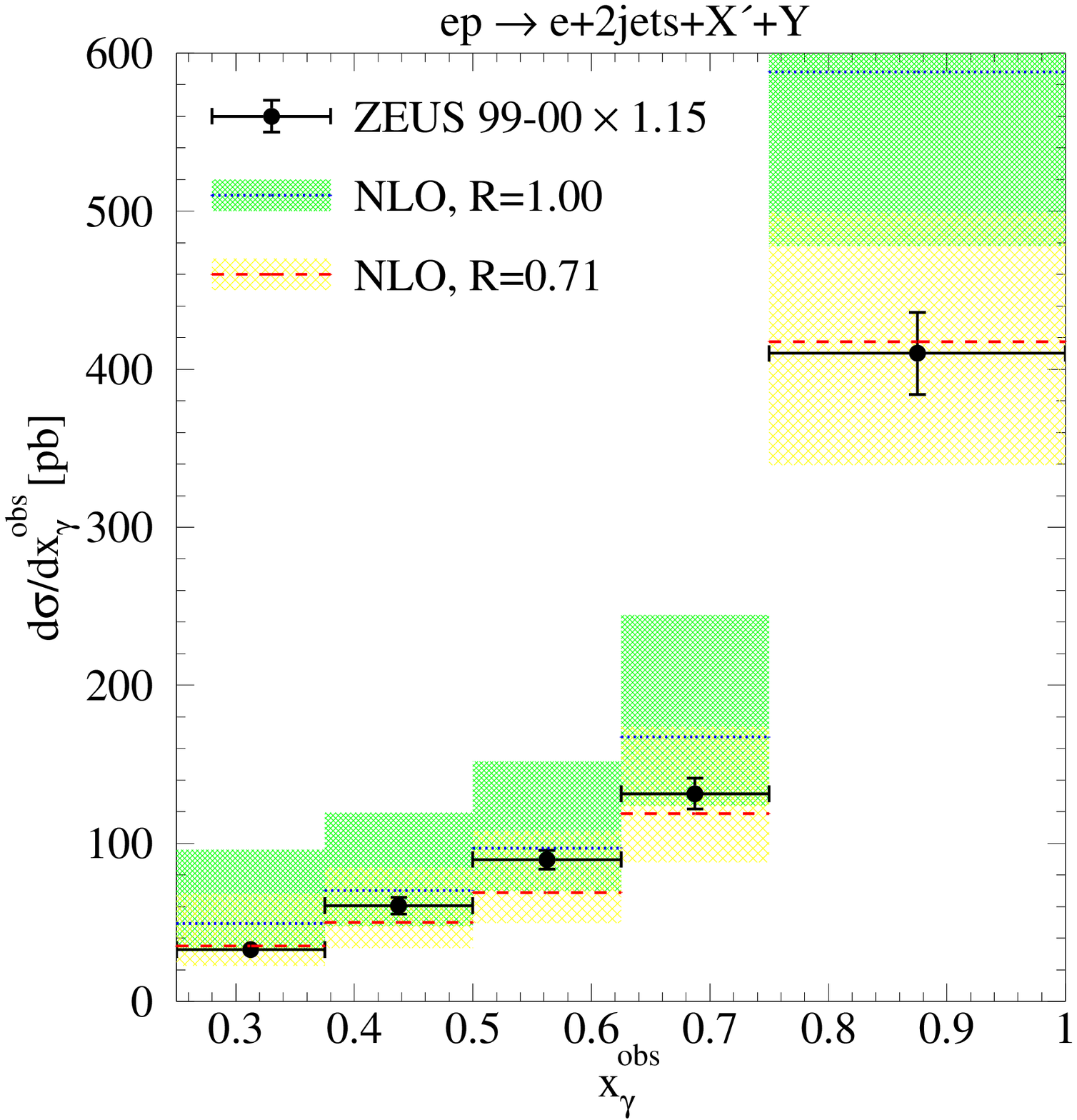}
 \includegraphics[width=0.325\columnwidth]{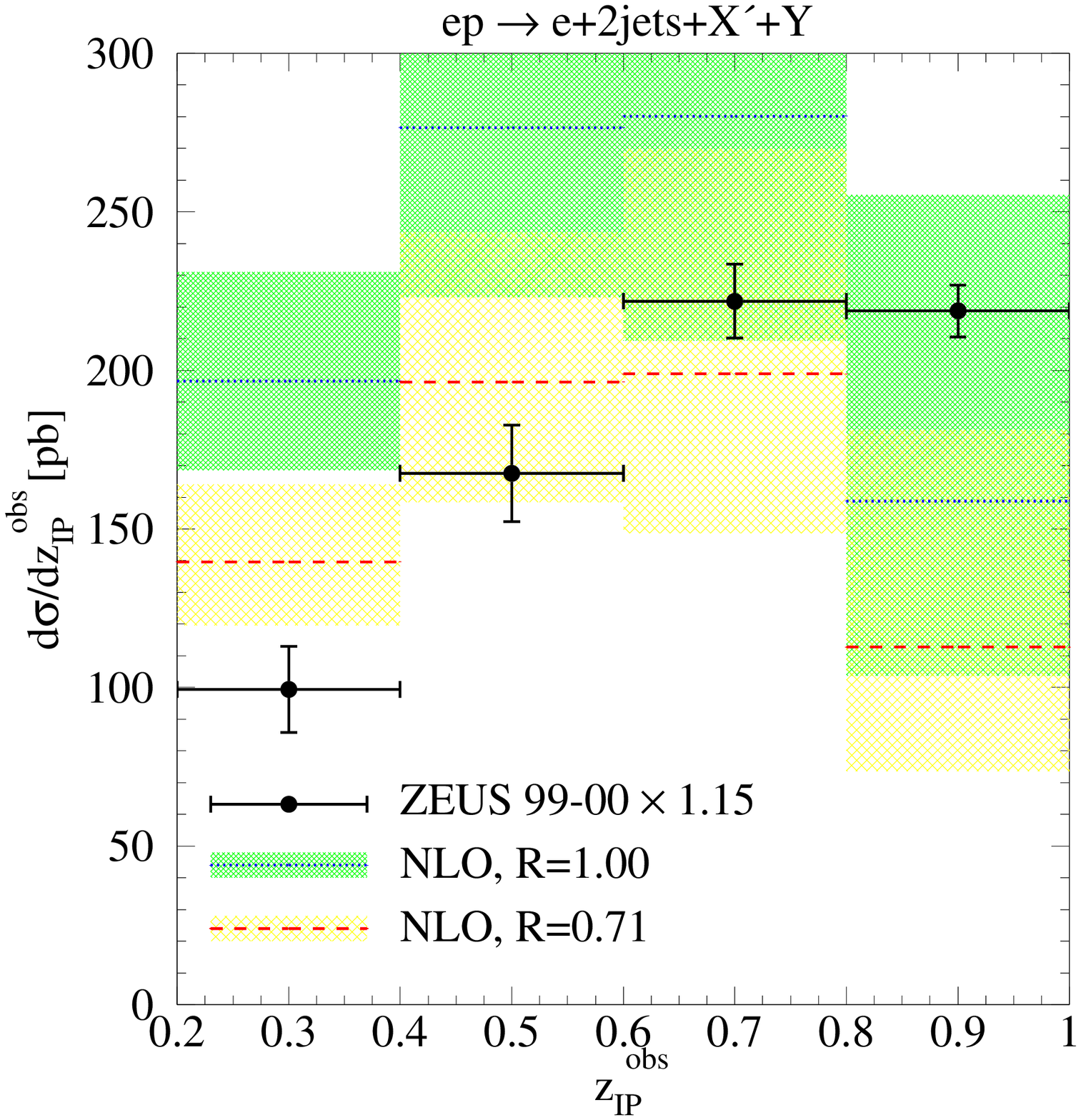}
 \caption{\label{fig:9}Differential cross sections for diffractive dijet
 photoproduction as measured by ZEUS and compared to
 NLO QCD without ($R=1$) and with ($R=0.71$) global suppression 
 (color online).}
\end{figure}
%%%%%%%%%%%%%% End of figure 9 %%%%%%%%%%%%%%%%%%%%%%%%%%%%%%%%%%%%%%%%%
%
%
%%%%%%%%%%%%% Begin figure 10 %%%%%%%%%%%%%%%%%%%%%%%%%%%%%%%%%%%%%%%%%%
\begin{figure}
 \centering
 \includegraphics[width=0.325\columnwidth]{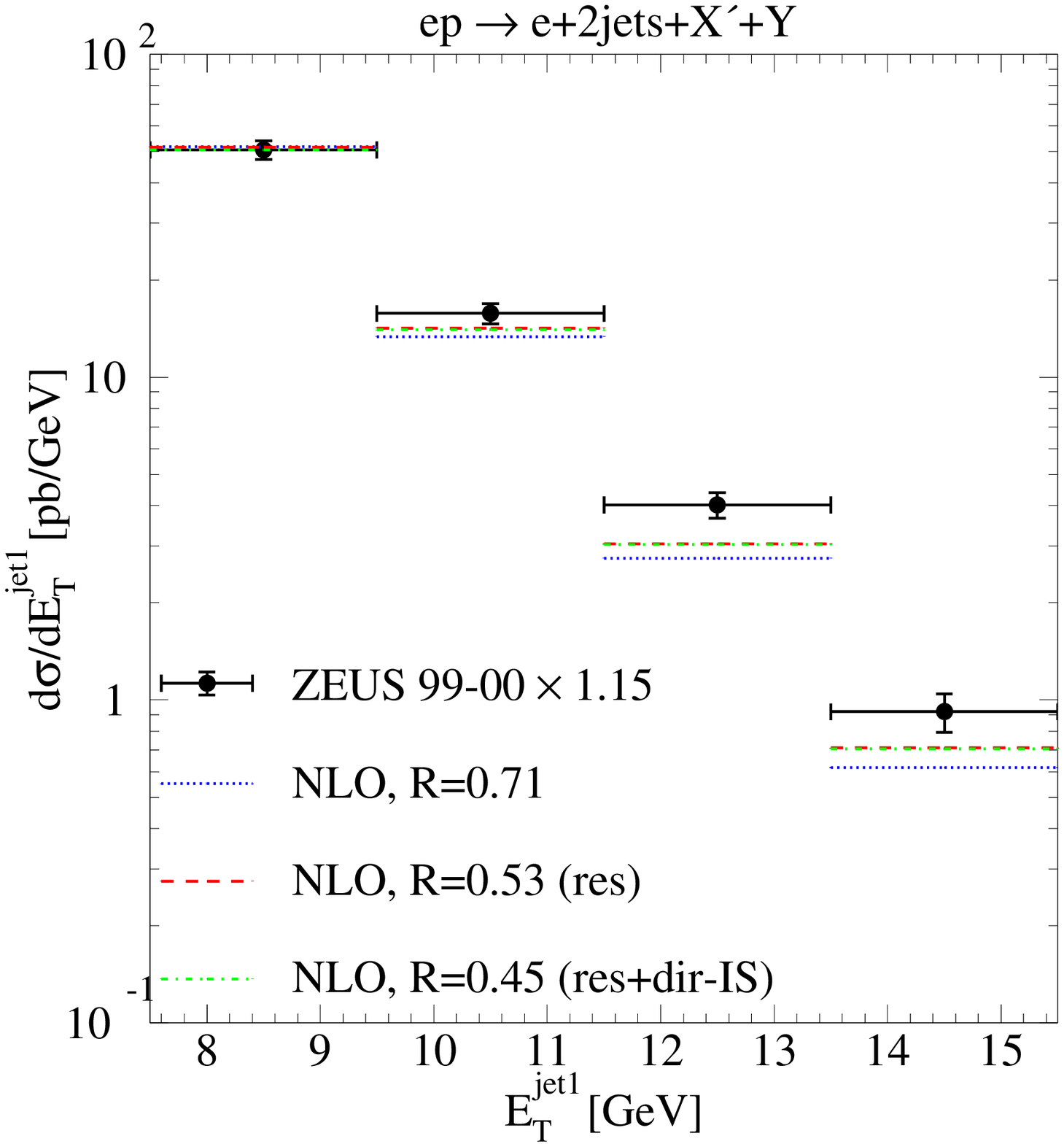}
 \includegraphics[width=0.325\columnwidth]{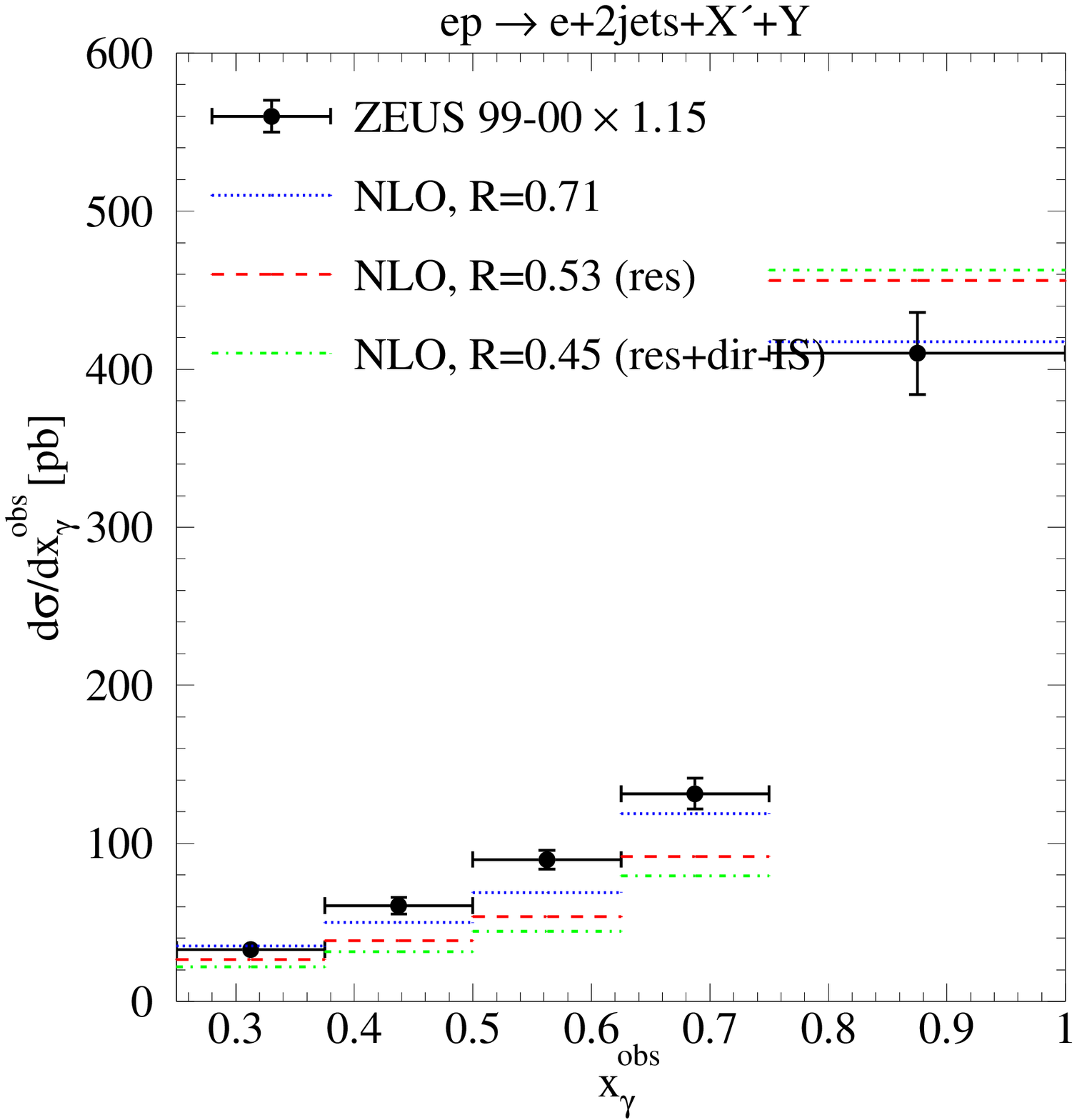}
 \includegraphics[width=0.325\columnwidth]{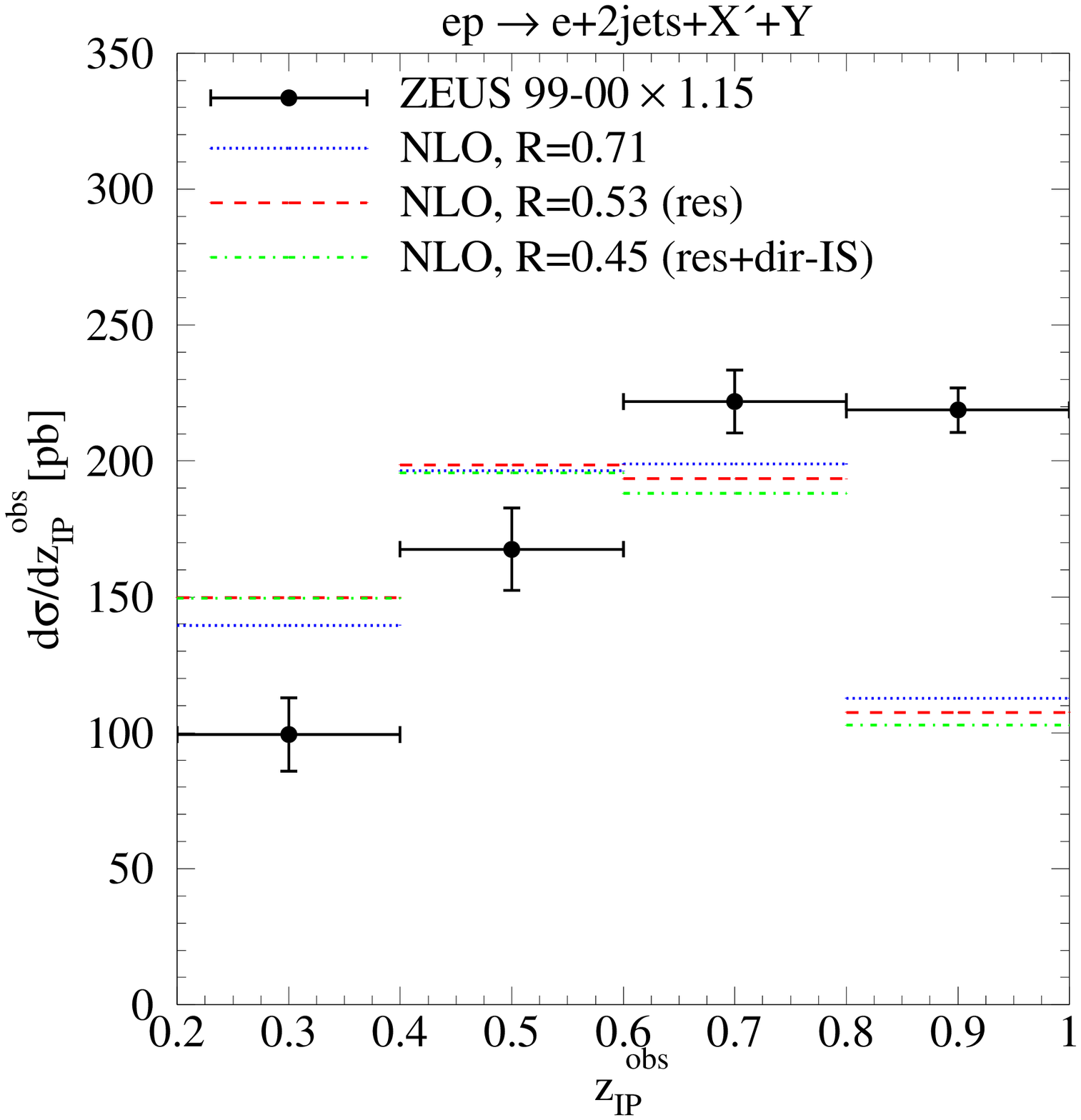}
 \caption{\label{fig:10}Differential cross sections for diffractive dijet
 photoproduction as measured by ZEUS and compared to
 NLO QCD with global, resolved, and resolved/direct-IS suppression.}
\end{figure}
%%%%%%%%%%%%% End of figure 10 %%%%%%%%%%%%%%%%%%%%%%%%%%%%%%%%%%%%%%%%%
%
In this section we shall compare our predictions with the final analysis
of the ZEUS data, which was published this year \cite{Chekanov:2007rh}, in order to see 
whether they are consistent with the large-$E_T^{jet}$ data of H1. The
kinematic cuts \cite{Chekanov:2007rh} are almost the same as in the
high-$E_T^{jet}$  H1 measurements. The only major difference to the H1 cuts 
%
%%%%%%%%%%%%%% Begin table 2 %%%%%%%%%%%%%%%%%%%%%%%%%%%%%%%%%%%%%%%%%%%
%\begin{table}
% \begin{tabular}{c}
% ZEUS cuts \\
% \hline
% $Q^2<$ 1 GeV$^2$  \\
% 0.2 $< y <$ 0.85     \\
% $E_T^{jet1} >$ 7.5 GeV \\
% $E_T^{jet2} >$ 6.5 GeV \\
% $-1.5 < \eta^{jet1(2)} < 1.5$ \\
% $x_{\p} <$ 0.025 \\
% $|t| <$ 5 GeV$^2$ \\
% \end{tabular}
% \caption{Kinematic cuts applied in the most recent ZEUS analysis of
% diffractive dijet photoproduction \cite{24}.}
%\end{table}
%%%%%%%%%%%%%% End of table 2 %%%%%%%%%%%%%%%%%%%%%%%%%%%%%%%%%%%%%%%%%%
%
is the larger range in the variable $y$.
Therefore the ZEUS cross sections
will be larger than the corresponding H1 cross sections.
% The different cuts
%on $Q^2$ and $|t|$  have little influence. For example, the larger $|t|$-cut 
%in Tab.\ 2 as compared to Tab.\ 1 increases the cross section only by $0.2\%$.
The constraint on $M_Y$ is not explicitly given in the ZEUS publication 
\cite{Chekanov:2007rh}. They give the cross section for the case that the diffractive final 
state consists only of the proton. For this they correct their measured cross
section by subtracting in all bins the estimated contribution of a
proton-dissociative background of $16\%$. When comparing to the theoretical
predictions they multiply the cross section with the
factor $0.87$ in order to correct for the proton-dissociative contributions,
which are contained in the DPDFs `H1 2006 fit A' and `H1 2006 fit B' by
requiring $M_Y < 1.6$ GeV. We do not follow this procedure. Instead we leave 
the theoretical cross sections unchanged, i.e.\ they contain a
proton-dissociative contribution with $M_Y < 1.6$ GeV and multiply the ZEUS 
cross sections by $1.15$ to include the proton-dissociative contribution. This 
means that the so multiplied ZEUS cross sections have the same proton
dissociative contribution as is in the DPDF fits of H1 \cite{Aktas:2006hy}. 
Since the ZEUS collaboration did measurements only for the high-$E_T^{jet}$
cuts, $E_T^{jet1(2)} > 7.5$ (6.5) GeV,
we can only compare to those. In this comparison 
we shall follow the same strategy as before. We first compared to the 
predictions with no suppression ($R = 1$) and then determine a suppression
factor by fitting $d\sigma/dE_T^{jet1}$ to the smallest $E_T^{jet1}$-bin.
Then we compared to the cross sections as a function of the seven observables 
$x_{\gamma}^{obs}$, $z_{\p}^{obs}$, $x_{\p}$, $E_T^{jet1}$, $y$, $M_X$ and
$\eta^{jet1}$
instead of the nine variables in the H1 analysis. The distribution in $y$ is 
equivalent to the $W$-distribution in \cite{h1dis08}. The theoretical predictions
for these differential cross sections with no suppression factor ($R = 1$) are
shown in Figs.\ 6a-g of \cite{Klasen:2008ah}, together with their scale errors and
compared to the ZEUS data points, and a selection is shown in Fig.\ \ref{fig:9}.
Except for the $x_{\gamma}^{obs}$- and
$E_T^{jet1}$-distributions, most of the data points lie outside the
theoretical error bands for $R = 1$. In particular, in Figs.\ 6b, c, e, f and
g, most of the points lie outside. This means that most of the data
points disagree with the unsuppressed prediction. Next, we determine the
suppression factor from the measured $d\sigma/dE_T^{jet1}$ at the lowest 
$E_T^{jet1}$-bin, 7.5 GeV $<E_T^{jet1}<9.5$ GeV, and obtain $R = 0.71$. This 
factor is larger by a factor of $1.15$ than the suppression factor from the 
analysis of the high-$E_T^{jet}$ data from H1. Curiously, this factor is
exactly equal to 
the correction factor we had to apply to restore the dissociative proton 
contribution. Taking the total experimental error of $\pm7\%$ from the 
experimental cross section $d\sigma/dE_T^{jet1}$ in the first bin
into account, the ZEUS suppression factor is $0.71\pm0.05$ to be 
compared to $0.62\pm0.14$ in the H1 analysis \cite{h1dis08}, so that both 
suppression factors agree inside the experimental errors.

If we now check how the predictions for $R = 0.71$ compare to the data
points inside the theoretical errors, we observe from Figs.\ 6a-g of Ref. 
\cite{Klasen:2008ah} that, with the exception of $d\sigma/dz_{\p}^{obs}$ and 
$d\sigma/dE_T^{jet1}$, most of the data points
agree with the predictions. This is quite consistent with the H1 analysis
(see above) and leads to the conclusion that
also the ZEUS data agree much better with the suppressed predictions than
with the unsuppressed prediction. In particular, the global suppression
factor agrees with the global suppression factor obtained from the
analysis of the H1 data inside the experimental error.

Similarly as in the previous section we compared the ZEUS data also with the
assumption that the suppression results only from the resolved cross
section. Here, we consider again (i) only resolved
suppression (res) and (ii) resolved plus direct suppression of the
initial-state singular part (res+dir-IS). For these two models we obtain the
suppression factors $R = 0.53$ and $R = 0.45$, respectively, where these
suppression factors are again obtained by fitting the data point at the first 
bin of $d\sigma/dE_T^{jet1}$. The comparison to the global suppression with 
$R=0.71$ and to the data is shown in Figs.\ 7a-g of \cite{Klasen:2008ah} and
a selection in Fig.\ \ref{fig:10}.
In general, we observe that the difference between global suppression and 
resolved suppression is small, i.e.\ the data points agree with the resolved 
suppression as well as with the global suppression. 

%In Figs.\ 11a and b the difference between `H1 2006 fit B' and `H1 2006 fit A'
%%
%%%%%%%%%%%%%% Begin figure 11 %%%%%%%%%%%%%%%%%%%%%%%%%%%%%%%%%%%%%%%%%
%\begin{figure}
% \centering
% \includegraphics[width=0.495\columnwidth]{fig10a}
% \includegraphics[width=0.495\columnwidth]{fig10b}
% \caption{\label{fig:11}Differential cross sections for diffractive dijet
% photoproduction as measured by ZEUS and compared to
% NLO QCD with resolved suppression and two different DPDFs.}
%\end{figure}
%%%%%%%%%%%%%%% End of figure 11 %%%%%%%%%%%%%%%%%%%%%%%%%%%%%%%%%%%%%%%%
%
%is shown again for the case of the resolved suppression. In both figures we
%observe that the fit A suppression with the suppression factor $R = 0.27$
%agrees better with the data than with the factor $R = 0.53$ for the
%fit B suppression. In particular, for $d\sigma/dE_T^{jet1}$ the agreement 
%with the three data points is perfect (note the logarithmic scale).

\section{Conclusion}

In conclusion, we found that most of the data points of diffractive dijet
photoproduction in the latest H1 analyses with low- and high-$E_T^{jet}$ cuts
and in the final ZEUS analysis with the same high-$E_T^{jet}$ cuts disagree
with NLO QCD predictions within experimental and theoretical errors. When
global factorization breaking is assumed in both the direct and resolved
contributions, the resulting suppression factor would have to be
$E_T^{jet}$-dependent, although we see no theoretical motivation for this
assumption. Suppressing only the resolved or in addition the direct
initial-state singular contribution by about a factor of three, as motivated
by the proof of factorization in point-like photon-hadron scattering and predicted
by absorptive models \cite{Kaidalov:2003xf}, the agreement between theory and data
is at least as good as for global suppression, and no $E_T^{jet}$-dependence of
the survival probability is needed.

%------------------------------------------------------------------------------
%       Bibliography
%------------------------------------------------------------------------------
\bibliographystyle{heralhc} 
{\raggedright
\bibliography{paper}

\providecommand{\etal}{et al.\xspace}
\providecommand{\coll}{Coll.}
\catcode`\@=11
\def\@bibitem#1{%
\ifmc@bstsupport
  \mc@iftail{#1}%
    {;\newline\ignorespaces}%
    {\ifmc@first\else.\fi\orig@bibitem{#1}}
  \mc@firstfalse
\else
  \mc@iftail{#1}%
    {\ignorespaces}%
    {\orig@bibitem{#1}}%
\fi}%
\catcode`\@=12
\begin{mcbibliography}{10}

\bibitem{Collins:1997sr}
Collins, J.C.,
\newblock Phys. Rev.{} {\bf D57},~3051~(1998)\relax
\relax
\bibitem{Affolder:2000vb}
Affolder, A.A. {\em et al.},
\newblock Phys. Rev. Lett.{} {\bf 84},~5043~(2000)\relax
\relax
\bibitem{Klasen:2002xb}
Klasen, M.,
\newblock Rev. Mod. Phys.{} {\bf 74},~1221~(2002)\relax
\relax
\bibitem{Klasen:2005dq}
Klasen, M. and Kramer, G.,
\newblock J. Phys.{} {\bf G31},~1391~(2005)\relax
\relax
\bibitem{Bruni:2005eb}
Bruni, A., Klasen, M., Kramer, G., and Schaetzel, S.
\newblock Prepared for the Workshop on the Implications of HERA for LHC
  Physics, CERN, Geneva, Switzerland, 26-27 Mar 2004\relax
\relax
\bibitem{Klasen:2005cz}
Klasen, M. and Kramer, G.,
\newblock AIP Conf. Proc.{} {\bf 792},~444~(2005)\relax
\relax
\bibitem{Kaidalov:2003xf}
Kaidalov, A.B., Khoze, V.A., Martin, A.D., and Ryskin, M.G.,
\newblock Phys. Lett.{} {\bf B567},~61~(2003)\relax
\relax
\bibitem{Aktas:2006hy}
Aktas, A. {\em et al.},
\newblock Eur. Phys. J.{} {\bf C48},~715~(2006)\relax
\relax
\bibitem{Aktas:2007hn}
Aktas, A. {\em et al.},
\newblock Eur. Phys. J.{} {\bf C51},~549~(2007)\relax
\relax
\bibitem{Chekanov:2007rh}
Chekanov, S. {\em et al.},
\newblock Eur. Phys. J.{} {\bf C55},~177~(2008)\relax
\relax
\bibitem{Klasen:2008ah}
Klasen, M. and Kramer, G.,
\newblock Mod. Phys. Lett.{} {\bf A23},~1885~(2008)\relax
\relax
\bibitem{Klasen:2004tza}
Klasen, M. and Kramer, G.
\newblock Prepared for 12th Int. Workshop on Deep Inelastic Scattering, Strbske
  Pleso, Slovakia, 14-18 Apr 2004\relax
\relax
\bibitem{Klasen:2004qr}
Klasen, M. and Kramer, G.,
\newblock Eur. Phys. J.{} {\bf C38},~93~(2004)\relax
\relax
\bibitem{h1dis08}
Aktas, A. {\em et al.}
\newblock Prepared for 16th Int. Workshop on Deep Inelastic Scattering, London,
  England, 7-11 Apr 2008\relax
\relax
\bibitem{kk08tbp}
Klasen, M. and Kramer, G.
\newblock DESY 08-109, LPSC 08-113, to be published\relax
\relax
\bibitem{Ellis:1993tq}
Ellis, S.D. and Soper, D.E.,
\newblock Phys. Rev.{} {\bf D48},~3160~(1993)\relax
\relax
\bibitem{Catani:1993hr}
Catani, S., Dokshitzer, Y.L., Seymour, M.H., and Webber, B. R.,
\newblock Nucl. Phys.{} {\bf B406},~187~(1993)\relax
\relax
\bibitem{Gluck:1991jc}
Glueck, M., Reya, E., and Vogt, A.,
\newblock Phys. Rev.{} {\bf D46},~1973~(1992)\relax
\relax
\bibitem{Aktas:2006qe}
Aktas, A. {\em et al.},
\newblock Phys. Lett.{} {\bf B639},~21~(2006)\relax
\relax
\end{mcbibliography}
}
\end{document}